\documentclass[12pt]{article}
\usepackage{epsfig}
\usepackage{epsf}
\usepackage{latexsym}
\usepackage{caption}
\begin{document}
\baselineskip 18pt
\def\simleq{\; \raise0.3ex\hbox{$<$\kern-0.75em\raise-1.1ex\hbox{$\sim$}}\; }
\def\simgeq{\; \raise0.3ex\hbox{$>$\kern-0.75em\raise-1.1ex\hbox{$\sim$}}\; }
\newcommand{\Dirac}{/\!\!\!\!D}
\newcommand{\GeV}{\,\mbox{GeV}}
\newcommand{\beq}{\begin{equation}}
\newcommand{\eeq}[1]{\label{#1}\end{equation}}
\newcommand{\bea}{\begin{eqnarray}}
\newcommand{\eea}[1]{\label{#1}\end{eqnarray}}
\newcommand{\Dslash}{\ensuremath \raisebox{0.025cm}{\slash}\hspace{-0.32cm} D}
\renewcommand{\Re}{\mbox{Re}\,}
\renewcommand{\Im}{\mbox{Im}\,}
\begin{titlepage}
\hfill  hep-th/0503247
\begin{center}
\hfill
\vskip .4in
{\Large\bf Naturalness in Cosmological Initial Conditions}
\end{center}
\vskip .4in
\begin{center}
{\large F. Nitti$^a$, M. Porrati$^{a,b}$ and J.-W. Rombouts$^a$}
\vskip .1in
{$^a$ \em Department of Physics, NYU, 4 Washington Pl.,
New York, NY 10003, USA}
\vskip .1in
{$^b$ \em Scuola Normale Superiore, 
Piazza dei Cavalieri 7, I-56126, Pisa, Italy}
\end{center}
\vskip .4in
\begin{center} {\bf ABSTRACT} \end{center}
\begin{quotation}
\noindent
We propose a novel approach to the problem of constraining
cosmological initial conditions.
Within the framework of effective field theory, we classify
initial conditions in terms of boundary terms added to the effective action 
describing the cosmological evolution below Planckian energies. These boundary
terms can be thought of 
as spacelike branes which may support extra instantaneous
degrees of freedom and extra operators. 
Interactions and renormalization of these boundary terms allow us to apply to
the boundary terms the field-theoretical requirement of naturalness, i.e. 
stability under radiative corrections. We apply this requirement to slow-roll 
inflation with non-adiabatic initial conditions, 
and to cyclic cosmology. This allows us to define in
a precise sense when some of these models are fine-tuned. 
We also describe how to parametrize in a model-independent way 
non-Gaussian initial conditions; we show that in some cases they are both
potentially observable and pass our naturalness requirement.
\end{quotation}
\vfill
 \hrule width 5.cm
\vskip 2.mm
{\small
\noindent e-mail: francesco.nitti@physics.nyu.edu, massimo.porrati@nyu.edu, 
jan-willem.rombouts@physics.nyu.edu}
\end{titlepage}
\tableofcontents
\vspace{1cm}
\begin{verse}
{\em Pronaque cum spectent animalia cetera terram,\\ os homini sublime dedit, 
caelumque videre\\ iussit et erectos ad sidera tollere vultus.}\\
Ovid, Metamorphoses, I 84-86
\end{verse}
\section{Introduction}
This old privilege notwithstanding, it took a long time to transform our
gazing into the sky into physics. The 
study of the early universe truly left the mist of myth and speculation to 
become science only in the 1930's, with the discovery of 
cosmic expansion. We had to wait until the 1960's, 
with the discovery of the cosmic microwave 
background (CMB), to be able to discriminate between the Hot Big Bang and 
alternative cosmologies, and only in the early 1990's, with the detection of
CMB's inhomogeneities, did cosmology fully become a quantitative science.

The celebrated WMAP survey~\cite{wmap} has spectacularly confirmed some
general predictions of slow-roll inflation, and offered the possibility of
significantly constraining alternative explanations for the primordial 
power spectrum. Future experiments may be able to go beyond the power spectrum
and check for other features of the CMB, as, for instance, primordial
non-Gaussianities.

Present data already make it meaningful to ask about finer details of the 
mechanism that generates an almost scale-invariant power spectrum. We just 
mentioned one such detail: possible non-Gaussian features. Another one is 
whether the correct initial state of the universe is the standard adiabatic 
``Bunch-Davies'' vacuum~\cite{bd}. 
Deviations from the standard inflationary vacuum offer the exciting 
possibility of observing 
very high-energy, ``trans-Planckian'' physics in the cosmic
microwave background radiation, thanks to the enormous stretch in proper
distance due to inflation~\cite{bm}. This effect, which may present 
us with a real chance for probing
string theory or any other model of quantum gravity, has received
considerable attention, once the possibility was raised that these
effects could be as large as $H/M$, with $H$ the Hubble parameter during
inflation, and $M$ the scale of new physics (e.g. the string scale).

Due to our ignorance of the ultimate theory governing high-energy physics,
the most natural, model-independent approach to studying modifications to the
primordial power spectrum is effective field theory (EFT)~\cite{kkls,bch}.
Using an EFT approach~\cite{kkls} concluded that
the signature of any trans-Planckian modification of the standard inflationary
power spectrum is $O(H^2/M^2)$, well beyond the reach of observation even in
the most favorable scenario ($H\sim 10^{14}\,\mbox{GeV}, M\sim 10^{16}\,
\mbox{GeV}$).
 
What was absent from e.g.
ref.~\cite{kkls} was a systematic EFT approach to
initial conditions. Ref.~\cite{kkls} presented convincing arguments against the
(in)famous $\alpha$-vacua~\cite{alpha} of de Sitter space, but it did
not give a complete parametrization of finite-energy, non-thermal states.
That parametrization was given in~\cite{sspds,gsss}, where the EFT approach was
systematically extended to the choice of initial conditions.
Refs.~ \cite{sspds,gsss} conclude that changes in the initial conditions for 
inflation are under control and may give $O(H/M)$ corrections to the 
primordial power spectrum. According to~\cite{sspds,gsss}, 
these corrections are quite characteristic
of UV modifications to physics and can be distinguished from other corrections
arising instead from IR changes in the vacuum state.
 
In this paper, we shall argue that other constraints on the EFT of 
initial conditions make $O(H/M)$ changes in the primordial power spectrum
unnatural, in the same sense that a light Fermi scale is unnatural in the
Standard Model. Before explaining further this point, we need to sketch the
approach of refs.~\cite{sspds,gsss}.  The most important difference 
between that and other approaches is that in~\cite{sspds,gsss}
initial conditions for modes of any wavelength are specified at the
same initial time $t^*$. Other approaches give the initial conditions
separately for each mode, at the
time it crosses the horizon. The latter prescription is useful
in the context of inflationary cosmology, but it obscures the field-theoretical
meaning of the perturbation and/or initial condition: it does not easily
account for the fact that after $t^*$ curvatures and energy densities 
are small,
so the field theory is under control, and it does not easily
translate into an EFT language. The former prescription, instead, leads
naturally to a simple classification of initial conditions in terms of
{\em local} operators defined at the space-like boundary
(i.e. initial surface) $t=t^*$. It also allows one to rephrase the question of
naturalness of initial conditions for inflation in terms of the usual 
field-theoretical notion of naturalness. 

In field theory we have a naturalness problem whenever the UV cutoff of the
theory, $M$ is much bigger than the observed value of the coefficients of
unprotected relevant operators. For instance, a quadratically-divergent
scalar mass term $m^2\phi^2$ is unnatural (or, equivalently, fine tuned) 
whenever $m \ll M$.
In our EFT theory of initial conditions, we shall define 
a parallel notion. Initial
conditions will be deemed unnatural whenever they require to fine-tune 
coefficients of relevant boundary operators to values much smaller than 
$M^{3-\Delta}$, where $\Delta$ is the dimension of the operator.

\begin{figure}[h]
\centering \epsfxsize=5in \hspace*{0in}\vspace*{.2in}
\epsffile{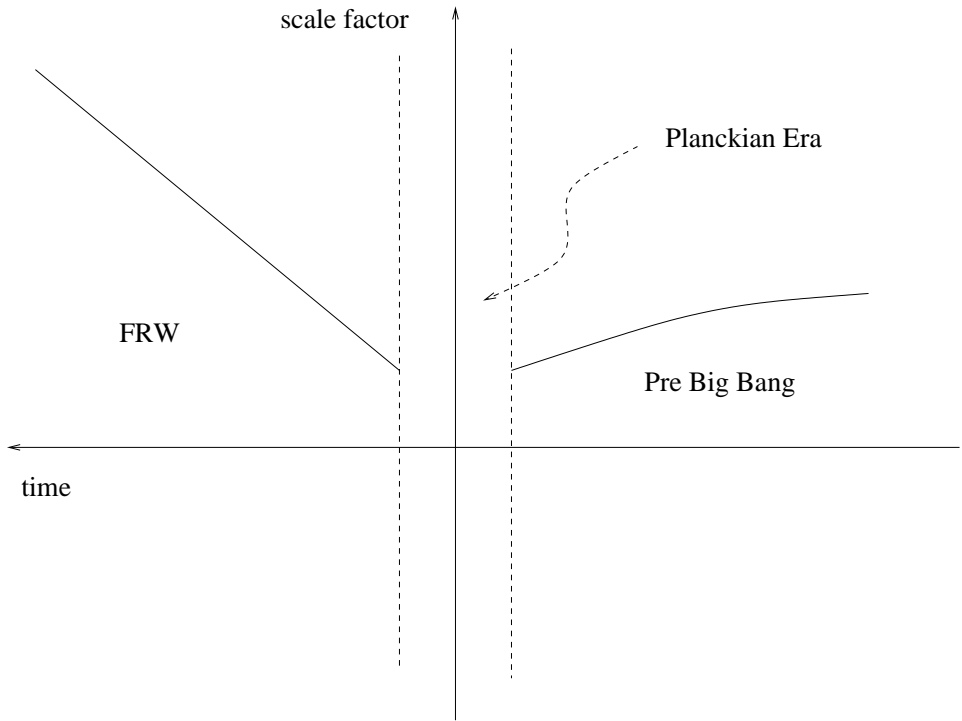} 
\caption{\small A schematic view of the EFT description of a cyclic cosmology. 
It plots the scale factor vs. time. 
The Planckian era around the bounce 
is replaced by an effective space-like brane (S-brane). } 
\label{bounce}
\end{figure}

Section 2 is devoted to summarize the EFT approach of~\cite{sspds,gsss}. 
There, we also generalize their formalism
to the case where 3-d boundary interactions are not added at an ``initial'' 
time, but they are used instead to parametrize an unknown period in 
the history of the universe. We shall do that to apply EFT methods to 
ekpyrotic/cyclic~\cite{ek} or pre-big bang cosmologies~\cite{gv}. In a 
sentence: we shall replace the unknown physics at the bounce --where the scale
factor shrinks to zero to originate a spacelike singularity-- with a 
spacelike (instantaneous) brane, whose world-volume supports local 
operators and additional degrees of freedom (see fig.~\ref{bounce}). 

Section 3 explains in greater details the concept of naturalness for initial
conditions. Naturalness is then applied to constrain possible changes in 
the B.D. vacuum of inflation. They turn out to be either unnatural or IR 
universal; therefore, ill-suited to characterize signals of new high-energy 
physics.

In section 4, we show that the spacelike brane
that parametrizes the unknown Planckian physics at the bounce in 
cyclic/ekpyrotic/pre-big bang  cosmologies can give raise to the correct power 
spectrum $P\sim k^{-3}$, generically at the price of severe 
non-localities on the brane. We show that typically these non-localities lead 
after the bounce to faster-than light propagation of signals originating in the
pre-big bang phase. This pathology may not be lethal for these models, since
Lorentz invariance is explicitly broken by the brane. Nevertheless, it is 
troublesome. It is avoided with a particular choice of interaction terms on
the brane, which we proceed to
show is unnatural according to our general prescription, 
since  it requires the fine-tuning to zero of an unprotected relevant
operator. 
The rest of section 4 is used to show how to 
write in local form a non-local  boundary
interaction on the brane, which can generate the correct post-bounce power 
spectrum. 
We show that the boundary term can be written in terms of local 
operators, but only at the price of introducing 
extra auxiliary fields that propagate
only on the spacelike brane. We then conclude the section by showing that,
besides leading to faster-than-light propagation, a generic boundary term  
also requires to fine-tune certain relevant operators, and it is thus also
unnatural. Our argument confirms and complements the analysis of~\cite{cnz}
(see also~\cite{bv}).

Section 5 shows how to compute the effect of changes in the initial state of
inflation that do not affect the power spectrum, but that do change the
three-point function of scalar fluctuations. 
We show that the changes induced by cubic boundary 
operators can lead to observable non-Gaussianities, different from those 
studied in~\cite{m}, without requiring undue fine-tunings.

Section 6 studies the three point function of scalar fluctuations in cyclic
cosmologies. Setting aside naturalness considerations, we show that, in this
scenario, any pre-big bang non-Gaussianity is damped. 
Finally, we show that the particular form of the scalar potential used in 
cyclic cosmologies sets a bound on how close to the singularity is the 
space-like brane. Specifically, if we want to avoid uncontrollably large 
interactions, the pre-big bang cosmic evolution must be cut off and replaced by
an effective brane at a time parametrically larger than $M_p^{-1}$. 

In section 7 we 
summarize our findings and conclude pointing out to possible developments of
our formalism to other problems involving spacelike singularities.
\section{EFT of Cosmological Initial Conditions}
\subsection{Inflation}
Differently from other approaches, here as in~\cite{sspds,gsss} we give 
initial conditions for modes of all wavelengths at the 
same initial time $t^*$. 

The prescription starts by supplementing the EFT action describing all 
relevant low energy fields with a boundary term that encodes the standard 
thermal vacuum. To be concrete, we begin by working out the example of a 
massless scalar field in a time-dependent background. 
The 4d (bulk) action plus a 3d boundary term is
\bea
S&=&S_4+S_3, \qquad S_4={1\over 2}\int_{t^*}^\infty dt \int d^3 x 
\sqrt{-g} g^{\mu\nu}\partial_\mu \chi^* \partial_\nu \chi , \nonumber \\
S_3 &=& \left. {1\over 2}\int d^3 x \sqrt{h^*(x)} 
\int d^3 y \sqrt{h^*(y)} \chi^*(x) \kappa(x,y) \chi(y)\right|_{t^*}.
\eea{q1}
Here $h^*_{ij}$ is the induced metric on the surface $t=t^*$. The role of
$S_3$ is to specify the wave functional for the scalar $\chi$ at $t=t^*$:
\beq
\Psi[\chi(x)]=\exp(iS_3[\chi]).
\eeq{q2}
Selecting an initial state for $\chi$ means in this language to choose a 
particular $\kappa(x,y)$. For instance, in de Sitter space with line element
\beq
ds^2= a(\eta)^2(-d\eta^2 + dx^idx^i), \;\;\; a(\eta)=-{1\over H\eta},
\;\;\; i=1,2,3, \;\;\; -\infty <
\eta <0,
\eeq{q2a} 
the standard thermal~\cite{bd} vacuum is obtained by choosing
\beq
\tilde{\kappa}(k)= -{|k|^2\eta^*\over 1+i|k|\eta^*}.
\eeq{q3}
Here, a tilde denotes the Fourier transform from space coordinates to 
co-moving momenta $k$ ($|k|\equiv\sqrt{k\cdot k}$) and $\eta^*$ is the initial 
(conformal) time.\footnote{From now on, $\eta$, $\eta^*$ will denote the
conformal time, $t$, $t^*$ will denote the synchronous proper time, and
$a()$ will always denote the scale factor. Notice that $\eta^*$ here 
is negative.}
This expression for $\kappa$ makes it clear that the choice of
such initial time is conventional, since a change in $\eta^*$ changes only 
$\kappa$, not the wave functional. From now on, whenever needed, 
the standard vacuum functional will be called $|0\rangle$.
\subsection{Changing the Initial State}
Next, we want to find a convenient classification of changes to the initial 
state. This can be done by adding a new boundary term to the action:
$S\rightarrow S+\Delta S_3$. To determine $\Delta S_3$, we notice that, at any
finite time $t$ after $t^*$, we are insensitive to changes that only affect
very low
co-moving momenta $k$: co-moving momenta $|k|< H(t) a(t)$ correspond to
perturbations with superhorizon physical wavelength $\lambda_p>1/H(t)$, which
are unobservable at time $t$. So, since we are interested in changes that can 
be observed in the CMB of the present epoch, we have an IR cutoff naturally
built into the theory. This IR cutoff tells us that observable changes in 
the initial conditions can be parametrized by {\em local} operators:
\beq
\Delta S_3 = \sum_i \beta_i  M^{3-\Delta_i} 
\left.\int d^3x \sqrt{h^*} O^i\right|_{t^*}.
\eeq{q4}
Here $O^i$ are operators of scaling 
dimension $\Delta_i$, $M$ is the high-energy 
cutoff of the EFT and the $\beta_i$'s are dimensionless parameters. 
The dimension $\Delta_i$ determines among other things how ``blue'' is the 
change in the power spectrum: the fractional change
in the power spectrum is proportional to $k^{\Delta_i-2}$. Since the
EFT makes sense only for $k<M$, operators of high conformal dimension do not 
significantly change the observable spectrum. So, the most significant 
observable changes in the primordial fluctuation spectrum are parametrized by 
a few local operators of low conformal dimension.

We just mentioned that the EFT needs a UV cutoff. This means that the operators
$O_i$ must be suitably regulated at short distance. In other words, they 
are local only up to the cutoff scale $M$. As a simple example, consider the 
dimension-four operator $O^4=(\beta/M)(\partial_i\chi)^2$. It has to be smeared
at short distance, for instance by the replacement
\beq
\partial_i\chi\partial_i\chi \rightarrow 
\partial_i\chi f(-\partial^2/a^2(t^*)M^2)\partial_i\chi.
\eeq{q5}
Here $f(x)$ is a smooth function obeying  $f(x)=1$, for $x\leq 1-\epsilon$;
$f(x)=0$, for $x\geq 1+\epsilon$; $\epsilon$ is a small positive number.
The scale factor 
$a(t^*)$ appears because we want to cutoff at $M$ the {\em physical} momentum 
$|k|/a(t^*)$, not the co-moving momentum $|k|$.
\subsection{Power Spectrum}
As a first application, let us derive the change in the power 
spectrum of a minimally-coupled scalar field in de Sitter space,
induced by the operator $O^4$ introduced in the previous 
subsection~\cite{sspds}. The change in initial 
conditions $\Delta S_3$, is equivalent to perturbing the Hamiltonian of the 
system by an instantaneous interaction $H_I=-\delta(\eta-\eta^*)\Delta S_3$. 
So, the perturbed two-point correlation function
is
\beq
G(k)=\lim_{\eta\rightarrow 0^-} \langle |\chi(k,\eta)|^2\rangle =
\lim_{\eta\rightarrow 0^-}
\langle 0| \exp (-i\Delta S_3) |\chi(k,\eta)|^2  \exp (i\Delta S_3)|0\rangle.
\eeq{q6}
To first order in $\beta$, the change is
\beq
\delta G(k) = -i {\beta\over M}\int d^3x a(\eta^*)\langle 0| [O^4(x), 
|\chi(k,0)|^2 ]|0\rangle.
\eeq{q7}
This quantity is easily computed in terms of commutators of free fields in de
Sitter space, resulting in~\cite{sspds,p1,p2}
\beq
\delta G(k) = -{\beta \over M}a(\eta^*) 
{H^2\over |k|^3} \Im [\chi^+(\eta^*,k)]^2 |k|^2
f(|k|^2/a^2(\eta^*)M^2).
\eeq{q8}
The canonically normalized, positive frequency solution of the 
free-field equations of motion is 
\beq
\chi^+(\eta,k)= {H\over \sqrt{2|k|^3}}(1+i|k|\eta)\exp(-i|k|\eta),
\eeq{q9}
and $ \langle 0||\chi(k,0)|^2|0\rangle = H^2/ |k|^3$ is the unperturbed two-point function.
For $|k||\eta^*| \sim 1$, the effect of $O^4$ on the power spectrum, $P(k)=(k/2\pi)^3G(k)$,  can be as
large as 
$\delta P/P \sim \beta H/M$, i.e. in the observable range when $\beta$ is 
$O(1)$. For $|k||\eta^*|> 1$, 
the oscillating exponential gives a characteristic
periodic signature~\cite{gsss}. The signal rapidly decays for
$|k||\eta^*|\gg 1$, so, it can be detected only if inflation lasts for a 
relatively short period; otherwise, $O^4$ would only affect unobservable 
super-horizon fluctuation.
\subsection{Scalar Metric Fluctuations}
Of course, we are not really interested in the fluctuations of a minimally 
coupled scalar. Those describe at best tensor perturbations of the metric. 
We want instead 
scalar perturbations of the metric. In the standard setting for
slow-roll inflation, where the field content is the metric $g_{\mu\nu}$ plus
the inflaton $\phi$. It is convenient to use the ADM formalism and decompose 
$g_{\mu\nu}$ into 3-d metric $h_{ij}$, shift $N^i$ and lapse $N$:
\beq
ds^2= -N^2dt^2 + h_{ij}(dx^i +N^i dt)(dx^j + N^j dt).
\eeq{m1}
Metric and scalar field split into background values plus fluctuations as
$h_{ij}=\exp(2\rho)(\delta_{ij}+ \gamma_{ij})$, $a(t)\equiv \exp(\rho)$,
$\psi=\phi(t)+\varphi$.
The last equation means that the background scalar field, $\phi(t)$, 
is a function of the time $t$ only. 
A particularly convenient gauge is~\cite{m}
\beq
h_{ij}= \exp(2\rho + 2\zeta) (\delta_{ij} + \gamma_{ij}), \qquad 
\partial_i \gamma_{ij}=0, \qquad \gamma_{ii}=0, \qquad \varphi=0.
\eeq{m2}
In this gauge, the scalar fluctuations are given by $\zeta$, which is also 
the gauge invariant variable of~\cite{b} (see also~\cite{bfm}). The quadratic
bulk action for $\zeta$ is almost identical with 
that of a massless scalar field
\bea
S&=&
{1\over 16\pi G}\int dtd^3x{\dot{\phi}^2\over \dot{\rho}^2}\left[ 
-e^{3\rho} \dot{\zeta}^2 + e^\rho (\partial\zeta)^2\right]. \label{m3}\\
&=& {1\over 16 \pi G}\int d\eta d^3x{\phi'^2\over \rho'^2}a^2(\eta)\left[ 
-\zeta'^2 + (\partial\zeta)^2\right].
\eea{m4}
For future reference, we have written the action both in synchronous time $t$ 
and in conformal time $\eta$. Derivative w.r.t. $t$ is denoted by a dot while
a prime denotes derivation w.r.t. $\eta$; $G$ is the Newton constant.
 
This formula shows that scalar fluctuations can be treated as a non-canonically
normalized scalar fields. So, the equations derived in the previous subsections
translate into formulas for changes in the CMB. One may worry that this 
conclusion has been reached with a very special choice of gauge. This is not 
the case, though. 
In the 4-d bulk, in any gauge, only the combination 
$\zeta - (\dot{\phi}/\dot{\rho})\varphi$ propagates. 
This combination is indeed the gauge-invariant definition of the Bardeen 
variable (see~\cite{b} and~\cite{bfm} section 10.3), and it is
this quantity that appears in the action.   
On the initial-time boundary, instead, both 
$\zeta$ and $\varphi$ appear~\cite{bfm}, but the latter appears only 
as a non-dynamical field, without a genuine kinetic term. Schematically:
\beq
S_{boundary}=\int d^3 x F[\zeta(x,t^*),\varphi(x,t^*)],
\eeq{m5}
where $F[\zeta,\varphi]$ is a quadratic function of its variables.
Therefore, after imposing the non-dynamical $\varphi$ equations of motion, 
$\partial F/\partial \varphi=0$,
$\varphi$ becomes a linear function of $\zeta$, that can be plugged back into
the boundary term to yield a quadratic function of $\zeta$ only.
\subsection{Cyclic Cosmologies}
We shall try to obtain an EFT description of cyclic cosmologies, independent 
of the details of their unknown high-curvature phase. In our description, the
period of cosmic evolution when perturbations are generated, 
separates into three periods (see fig. 1): a slow contraction,
where the scale factor evolves as $a(\eta)=(-\eta/\eta^*)^\varepsilon$, 
$\eta < -\eta^*$,
$\varepsilon \ll 1$\footnote{In the pre-big bang model of~\cite{gv}, instead, 
$\varepsilon=1/2$. Here $\eta^*$ is positive.};
a high-curvature bounce at $|\eta| \approx \eta^*$, the unknown dynamics 
whereof is replaced by a 
surface term (a spacelike brane, or S-brane); a FRW phase, typically a
radiation-dominated phase with $a(\eta)=\eta/\eta^*$, $\eta > \eta^*$.
The scalar $\phi$ is no longer a slow-roll inflaton. In the pre-bounce, 
pre-big bang phase, the potential is a very steep 
exponential over some range of $\phi$, for instance~\cite{kost}:
\beq
V=-V_0 \exp(-\sqrt{16\pi G /p} \,\phi), \qquad 0< p \ll 1,
\eeq{m6}
and the background is
\beq
a(\eta)=(-\eta/\eta^*)^{p/(1-p)}, \qquad \phi(\eta)=
{\sqrt{2p/8\pi G}\over 1-p} \log (-\eta/\eta^*),
\eeq{m7}
 where the parameters $V_0$ and $\eta^*$ are related by
 $\eta^*=\sqrt{p(1-3p) /8\pi G V_0}$.
After the bounce, the scalar decouples from the dynamics and one finds the 
action for scalar fluctuations in a radiation-dominated FRW cosmology.

To match the two phases, we need to add a boundary term, an S-brane, with 
nonzero tension. Its form is largely free; we can only say that it must
contain at least the following terms: 
\beq
T_+\int d^3x \sqrt{h^*_+} - T_-\int d^3x \sqrt{h^*_-} + ....
\eeq{m7a}
Here $T_+$ and $T_-$ are constant tensions, needed to satisfy the junction 
conditions at $\eta=\eta^*$ and $\eta=-\eta^*$, respectively. $h^*_\pm$ is
the determinant of the induced metric on the surface $\eta=\pm\eta^*$ and 
$...$ stands for gradient terms that vanish on the background $\zeta=0$.

The nonzero tensions generate a stress-energy tensor, which is conserved but
which violates the null energy condition (NEC). 
This is unavoidable if we want a spatially flat bounce.

Thanks to the S-brane, we can have a bounce and we can 
write the action for scalar fluctuations in the gauge~(\ref{m2}) as
\bea
16\pi G S&=& 
\int_{\eta^*}^{\infty} d\eta \int d^3x \left({\eta\over\eta^*}\right)^2\left[ 
-\zeta'^2 + (\partial\zeta)^2\right] +
 {2\over p}\int_{-\infty}^{-\eta^*} d\eta \int d^3x 
\left({-\eta\over \eta^*}\right)^{2p/(1-p)}\left[ 
-\zeta'^2 + (\partial\zeta)^2\right]   \nonumber \\ &+& \int d^3x \left\{
\Lambda[\zeta (\eta^*)-\alpha(\partial)\zeta(-\eta^*)] + 
\zeta(\eta^*)F(\partial)\zeta(\eta^*) \right\}.
\eea{m8}

This is one of the most important equations in our paper, and it is worth of a
few comments.

First of all, the boundary term parametrizing the unknown high-energy physics
at the bounce is made of two pieces. The first is the one containing the 
Lagrange multiplier $\Lambda(x)$. 
This is a non-dynamical field whose role is to
link the value of the field $\zeta$ after the bounce, at $\eta=\eta^*$, to its
value before the bounce, at $\eta=-\eta^*$. $\zeta$ can be continuous at the 
bounce, for $\alpha=1$, or it can jump, whenever $\alpha\neq 1$. 
$\alpha(\partial)$ is a local function of spatial gradients. So, 
$\zeta (\eta^*)$
is generically a function of $\zeta(-\eta^*)$ and its derivatives along the 
brane. Likewise, $F(\partial)$ is a function of spatial gradients. 
Generically, it is nonlocal. Its role is to mimic super-horizon correlations
induced by the unknown physics at the bounce. Superhorizon correlations do not
necessarily signal acausalities, because our parametrization can also fit a 
``long'' bounce, lasting for an arbitrarily long time.

Notice that we did not introduce $d\zeta/d\eta$ in the boundary term. The 
reason is that  $d\zeta/d\eta$ terms render the variation of eq.~(\ref{m8})
ill-defined. So, whenever they do appear in a boundary term, they should be 
eliminated by an appropriate discontinuous field redefinition~\cite{sspds}, or
by using the bulk equations of motion to convert them into functions of $\zeta$
and its spacelike gradients.

We wrote eq.~(\ref{m8}) in a specific gauge. To write it in a 3-d covariant 
form, we must re-express $\zeta$ in terms of the intrinsic curvature on 
surfaces defined by $\varphi=0$. To write it in a fully 4-d covariant manner, 
we must furthermore express the position of the bounce in terms of a covariant
(scalar) equation involving, say, $\psi$ and the scalar curvature. 

We may worry about two features of this procedure. 
\begin{enumerate}
\item 
A covariant equation for
the position of the bounce reduces, in the gauge~(\ref{m2}), to
\beq
\eta=G(\zeta)+\eta^*.
\eeq{m9}
The function $G$ is smooth and it vanishes at $\zeta=0$, but is otherwise 
unknown. So, generically, the bounce does not sit at a constant value of the
time $\eta=\eta^*$. This is not a problem since at quadratic order
this bending of the brane
only generates further quadratic terms of the form $\zeta F \zeta$, plus a 
finite renormalization of the brane tensions $T_\pm$.\footnote{The brane 
bending also introduces terms in $\zeta'$. As we have already
mentioned, these terms
must be canceled by field redefinitions, or expressed in terms of gradients
via the equations of motion.} Explicitly, the induced metric on the brane is
\beq
h_{ij}^*=\partial_i G \partial_j G g_{00}(G(x),x) + \partial_i G g_{0j}(G(x),x)
+ \partial_j G g_{0i}(G(x),x) + g_{ij}(G(x),x).
\eeq{m9a}
By substituting this formula into the universal term $T\int d^3x \sqrt{h^*}$ 
we get, at linear order in $\zeta$ 
\beq
T\int d^3 x a^3(3\zeta + 3\rho' G_\zeta \zeta), \qquad G_\zeta\equiv 
\left.{dG\over d\zeta}\right|_{\eta=G(x)}.
\eeq{m9b}
This term induces a finite renormalization of the brane tension,
$T\rightarrow T(1 +\rho' G_\zeta)$.
At quadratic order in $\zeta$ we find as announced a term of the form
$\zeta F \zeta$, after eliminating time derivatives by the bulk
equations of motion:
\bea
&& T\int d^3 x a^3\left[ G_\zeta ^2 (\partial_i \zeta)^2 + 2
G_\zeta \partial_i \zeta N_i a^{-2} + 3(2\rho^{'2} + \rho'' ) G_\zeta^2 \zeta^2
+6\zeta^2 + 6G_\zeta\zeta \zeta'  + \right.\nonumber \\ 
&& \left. 3\rho' G_{\zeta\zeta} \zeta^2 +{3\over 4}(2\zeta+
2\rho' G_\zeta \zeta)^2\right].
\eea{m10}
\item The functions $\alpha$ and $F$ we introduced, are quite arbitrary.
They can even break explicitly 3-d rotations and translations, even though we
will not do that in the following. So, neither the momentum constraint nor the
Hamiltonian constraint impose any restriction on them. This is because our 
``initial time'' brane is different from the boundary brane introduced in the
context of de Sitter holography in~\cite{lm}. For us the brane is real. It 
either parametrizes a very specific time in cosmic evolution (the bounce), or a
specific choice of quantum state at some time early in the 
inflationary epoch. Since we use the brane to specify an initial state, which 
is then used to compute expectation values of $\zeta$ at 
$\eta \rightarrow \infty$ (that is today), we can change the value of 
$\eta^*$, together with the form of the boundary terms, in such a way as 
to keep the late-time correlators constant. An example of this 
``renormalization group'' is our eq.~(\ref{q3}). So, our boundary terms at
$\eta^*$ are the analog of {\em arbitrary} initial conditions for the RG group.
In ref.~\cite{lm} instead, the late-time brane is a regulator, and the time 
evolution of all fields is fixed, because both late-time and early-time
boundary conditions are fixed. The difference between the two approaches is 
schematically shown in fig. 2.
\end{enumerate}
\begin{figure}[h]
\centering \epsfxsize=6.5in \hspace*{0in}\vspace*{.2in}
\epsffile{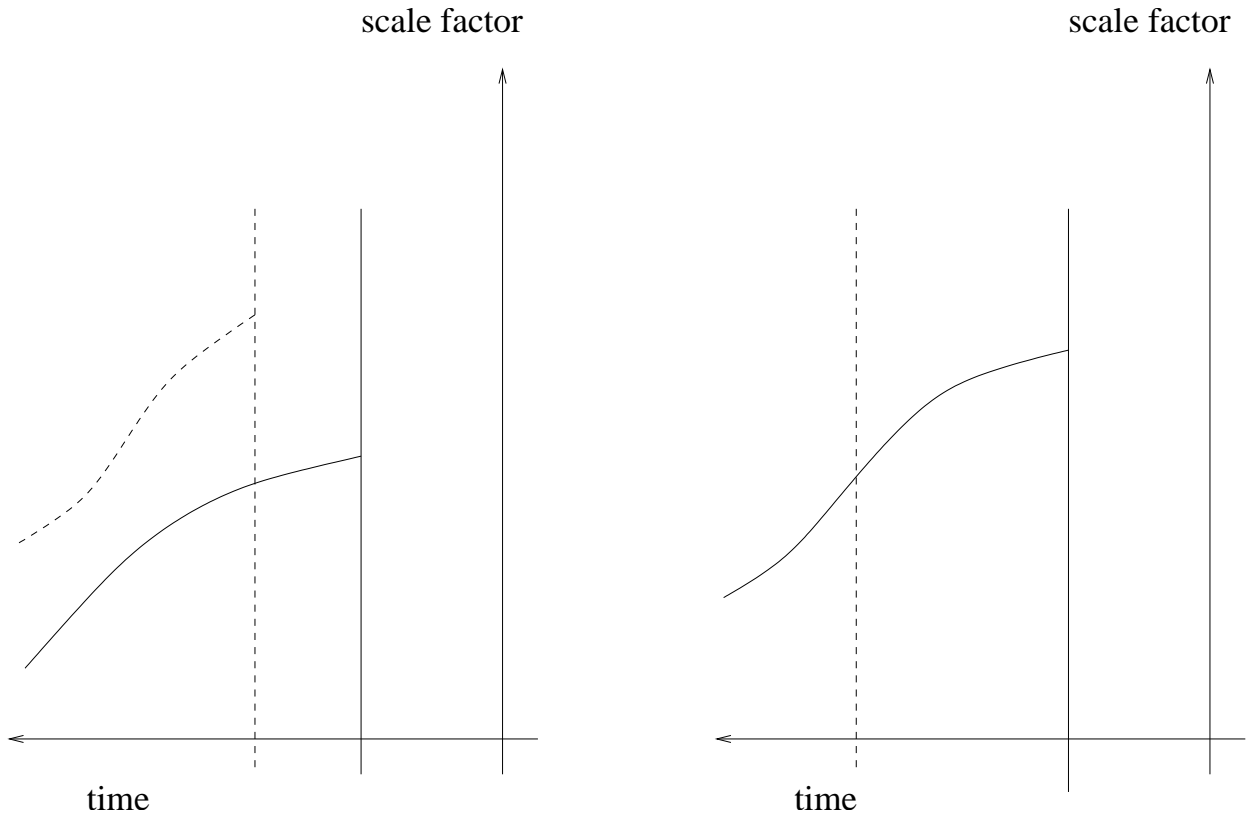}
\caption{\small On the left, a schematic view of our brane, its position is
represented by the vertical lines. By changing the 
operators on the brane and/or its position, we can 
change the initial conditions for the flow, hence the flow itself. 
On the left, the holographic brane
of ref.~\cite{lm}. Its boundary operators are constrained 
by requiring that the flow 
remains the same when its position is changed.}
\label{rgflow}
\end{figure}

\section{Naturalness}
In particle physics, stability under radiative corrections is a powerful 
guide for constraining high-energy extensions of the standard model or of any
EFT. By construction, an EFT has a built-in UV cutoff. In other words, the
EFT only describes the low-energy sector of a theory whose UV completion is
unknown. EFT is a powerful method whenever there is a large energy gap between
a known low-energy sector (say the standard model) and an unknown high-energy
sector --typically made of heavy particles-- 
that decouples below a cutoff $M$. In this case, the effect of 
integrating out the high-energy sector is to introduce irrelevant operators in 
the EFT. They have dimension $\Delta > 4$ and appear in the EFT 
with coefficients $O(M^{4-\Delta})$. A change in the UV physics --as for 
instance a change in masses and couplings of the heavy sector-- induces a 
small modification in the coefficients of these irrelevant operators. 
So, low-energy physics is shielded from changes in the UV physics, and EFT can
be predictive even in the absence of a complete knowledge of high-energy 
physics. The problem lies in the relevant operators; those with dimension 
$\Delta < 4$. Their coefficients must be much smaller than $M^{4-\Delta}$, 
otherwise, the EFT would describe a trivial physics with no light states at 
all. A simple example of this pathology is a theory with two scalars, one 
heavy, with  mass $M$, and another one with mass $m$. Below the energy
scale $M$ only one scalar propagates, and only if $m \ll M$. Relevant operators
are very sensitive to changes in the UV physics, 
the more so the lower their dimension. Generically, they appear in the EFT with
coefficients $O(M^{4-\Delta})$. Even if the tree-level values of these 
parameters are small, radiative corrections typically bring them to values
$O(M^{4-\Delta})$. Equivalently, any small change in masses, couplings etc. in
the heavy sector bring these coefficients to their typical value.

The best known naturalness problem in particle physics is that associated with
the Fermi scale $M_F\approx 100 \GeV$ or, equivalently, the Higgs mass.
Suppose that the standard model holds up to a high energy scale  
$M \approx 10^{16}-10^{19} \GeV$. Then, unless the Higgs mass is protected by  
symmetries (e.g. supersymmetry) it would be destabilized by radiative
corrections, and driven to the UV cutoff.

In our setting, we have a naturalness problem too. It arises because we change
the cosmological initial state by adding boundary terms to a bulk
action as in eq.~(\ref{q4}). These boundary terms are generically unstable 
under radiative corrections. So, even if we begin by modifying our theory 
by adding a (safe) irrelevant boundary operator\footnote{Since the boundary
is a 3-d field theory, all operators of dimension greater than 3 are 
irrelevant.}, we may end up generating dangerous relevant operators.
\subsection{Naturalness in Inflation I}  
We begin by studying in detail the example given in section 2.3. There we
computed the change induced 
in the fluctuation spectrum of a minimally-coupled massless scalar by the 
dimension-4 operator
\beq
O^4={\beta\over M}(\partial_i\chi)^2.
\eeq{m11}
This boundary term is not complete. It must be covariantized with respect to
3-d general coordinate transformations, which are not broken by a choice of
initial conditions. The covariantization is obvious and gives the following
boundary term
\beq
 \Delta S_3= {\beta\over M}
\int d^3x \sqrt{h}h^{ij}\partial_i\chi \partial_j\chi.
\eeq{m12}
Once we impose a shift symmetry $\chi \rightarrow \chi + \mbox{constant}$,
$O^4$ is the lowest-dimension boundary operator quadratic in $\chi$.
So, it is not unnatural to set to 
zero the coefficient of the relevant --and potentially dangerous-- operator
$O^2=\alpha M \chi^2$. 

In the gauge~(\ref{m2}), the boundary term~(\ref{m12}) gives rise to 
interaction terms between $\chi$ and $\zeta$. In particular, we get
a quartic interaction
\beq
 \Delta S_3^{\zeta^2\chi^2}= 
{\beta\over 2M}\int d^3x a(\eta^*)\zeta^2\partial_i\chi \partial_i\chi.
\eeq{m13}
This is a dangerous interaction. Its effect is best seen by writing the wave
function at late time as a functional integral
\beq
\Psi[\zeta_{now},\chi_{now}]=\int [d\zeta d\chi..] 
e^{i(S_4 + S_3 + \Delta S_3)}.
\eeq{m14}
$\chi,\zeta$ are given free boundary conditions at the initial time $\eta^*$,
while at late time, $\eta \rightarrow 0^-$, they are given fixed (Dirichlet)
boundary conditions
\beq
\zeta(\eta)|_{\eta=0}= \zeta_{now}, \qquad \chi(\eta)|_{\eta=0}=\chi_{now}.
\eeq{m15}
This functional integral can be computed perturbatively either in terms of 
Feynman diagrams or using the Hamiltonian formalism of subsection 2.3. 
Results become clearer using the Feynman diagram approach. 

First of all, bulk interactions may be potentially dangerous, since naively
the bulk action has interaction terms of the form 
$(16\pi G)^{-1} \zeta \partial\zeta\partial\zeta$, while the kinetic term is
multiplied by the extra factor $\epsilon=\phi'^2/\rho'^2$ [see 
eq.~(\ref{m4})], which is small in 
slow-roll inflation. Nevertheless, as shown in~\cite{m}, a field redefinition
of $\zeta$ eliminates all such terms and leaves only bulk terms of the form 
$(16\pi G)^{-1} \epsilon^2\zeta \partial\zeta\partial\zeta$. 

The boundary interaction~(\ref{m13}), instead, does produce a dangerous effect.
By integrating out $\chi$, this interaction generates a new boundary 
term for $\zeta$ thanks to the diagram in fig. 3.
 \begin{figure}[h]
\centering \epsfxsize=3in \hspace*{0in}\vspace*{.2in}
\epsffile{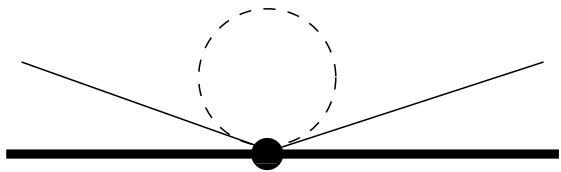} 
\caption{\small Self-energy correction to the $\zeta^2$ boundary ``mass'' 
terms. $\zeta$'s are denoted by solid lines and $\chi$'s by broken lines.} 
\label{selfenergy}
\end{figure}
\beq
\Delta S^{\zeta^2}= {\beta \over 2 M} \int d^3 x a(\eta^*)\zeta^2 
\langle \partial_i\chi \partial_i\chi \rangle.
\eeq{m16}
The propagator for $\chi$ is the standard one, computed in the
BD vacuum
\beq
\langle \chi(\eta,k)\chi(\eta',-k)\rangle 
= {H^2\over 2|k|^3}(1+i|k|\eta)(1-i|k|\eta')\exp[i|k|(\eta'-\eta)], 
\qquad \eta \geq \eta'.
\eeq{m17}
By substituting this expression, computed at $\eta=\eta^*$, 
into eq.~(\ref{m16}) we get 
\beq
\Delta S^{\zeta^2}\approx {\beta \over 2 M} \int d^3 x a(\eta^*)\zeta^2  
\int_{{|k|\over a(\eta^*)} \leq M} {d^3 k\over (2\pi)^3} 
{H^2\over 2|k|^3} |k|^2(1+|k|^2\eta^{*2})
\approx {\beta\over 96\pi^2}M^3\int d^3 x a^3(\eta^*)\zeta^2.
\eeq{m18}
Here we used a sharp momentum cutoff and the inequality $M\gg H$. Any other
cutoff would give a like result.

The effect of this term on the power spectrum is enormous, since the kinetic
term for scalar fluctuations is multiplied by the small number $\epsilon$.
If we canonically normalize $\zeta$ by 
\beq
\zeta=\sqrt{8\pi G \over \epsilon}v,
\eeq{m19}
we can use the same technique as in subsection 2.3 to get
\beq
{\delta P(k)\over P(k)} = C a^3(\eta^*)\Im [v^+(\eta^*,k)]^2,\qquad 
v^+(\eta,k)= {H\over \sqrt{2|k|^3}}(1+i|k|\eta)\exp(-i|k|\eta),
\qquad {|k|\over a(\eta^*)}\ll M.
\eeq{m20}
With our normalization, $C=\beta G M^3/ 12\pi \epsilon$. So, 
for $k\sim 1/\eta^*$, the power spectrum receives corrections
$O(\beta G M^3/ 12\pi H\epsilon)$. By imposing the observational constraint
$\delta P(k)/P(k) \simleq \epsilon$ we arrive at our first naturalness 
constraint on $\beta$
\beq
\beta \simleq  12\pi \epsilon^2 (GM^2)^{-1} {H \over M}.
\eeq{m21}
Notice that the power corrections we computed in section 2.3 were at most
$O(\beta H/M)$. Since the term~(\ref{m18}) is also linear in $\beta$ and as
large as  $\beta G M^3/ 12\pi H\epsilon$, it dominates over the tree-level
term whenever $G M^4/ 12\pi\epsilon H \simgeq 1$.  
Constraints of this kind  were derived with a different 
method in~\cite{p1,p2}. The method used here makes it clear that the bound
comes from asking that the correction to the BD initial state is generic and
robust under radiative corrections. Of course, as with any divergence, 
even that in eq.~(\ref{m18}) can be canceled by appropriately choosing 
boundary counter-terms. The point is that this choice implies a fine-tuned
UV completion of our EFT, which we should not assume without a valid reason 
such as a symmetry, or a better knowledge of the UV physics.

Notice that the counter-term that cancels~(\ref{m18}) is not the same that 
renormalizes the brane tension. The brane tension is renormalized because 
when we expand eq.~(\ref{m12}) in powers of $\zeta$, we also get a linear term
\beq
 \Delta S_3^{\zeta\chi^2}= 
{\beta\over M}\int d^3x a(\eta^*)\zeta\partial_i\chi \partial_i\chi.
\eeq{m22} 
This term can be canceled by changing the tension of the term
\beq
T\int d^3x \sqrt{h} =  T\int d^3x a^3(\eta^*) e^{3\zeta} =
T\int d^3x a^3(\eta^*)\left(1+ 3\zeta + {9\over 2} \zeta^2+..\right).
\eeq{m23}
Equation~(\ref{m18}) implies we can cancel the linear term~(\ref{m22}) by
changing the tension as 
\beq
3\delta T+ {\beta\over 48\pi^2}M^3 =0.
\eeq{m24}
To cancel the quadratic term, instead, we would need to change the tension
as
\beq
9\delta T+ {\beta\over 48\pi^2}M^3 =0.
\eeq{m25}
\subsection{Naturalness in Inflation II}
In the previous subsection, we introduced an extra scalar field, $\chi$, 
besides the inflaton. This was done to simplify our analysis 
and is by no means 
necessary. We could have introduced a boundary 
interaction involving only $\zeta$, for instance\footnote{The factor 
$\epsilon(8\pi G)^{-1}$ is included for convenience so that for the 
canonically normalized field $v$ the coefficient reduces to $\gamma/M$.} 
\beq
\Delta S_3={\gamma\over M}{\epsilon\over 8\pi G }\int d^3 x a(\eta^*)\zeta \partial^2 \zeta.
\eeq{m26}
This boundary term
 can be  covariantized by recalling that at linear order 
the 3-d Ricci curvature ${\cal R}_{ij}$ is proportional to
$\partial_i\partial_j\zeta$  plus
a term linear in the transverse traceless fluctuation $\gamma_{ij}$. 
So, the 3-d covariant form of eq.~(\ref{m26}) becomes
\beq
\Delta S_3={\gamma\over M}{\epsilon\over 8\pi G }\int d^3 x \sqrt{h} {\cal R}.
\eeq{m27}
Besides the quadratic term~(\ref{m26}), eq.~(\ref{m27}) produces, among others,
a quartic interaction term. In terms of the canonically normalized field
$v$ it reads
\beq
\Delta S_3^{v^4}={4\pi G\gamma\over M\epsilon}
\int d^3 x {1\over a(\eta^*)}
v^3 \partial^2 v.
\eeq{m28}
The self-energy loop now produces a boundary term of the form
\beq
{\gamma G M^3\over 9\pi\epsilon}\int d^3 x a^3(\eta^*)v^2.
\eeq{m29}
Following the same steps we used to arrive to eq.~(\ref{m21})
we obtain a constraint on $\gamma$  
\beq
\gamma \simleq  9\pi \epsilon^2 (GM^2)^{-1} {H\over M}.
\eeq{m30}

Covariantization of the boundary terms is just one of many ways in which dangerous boundary 
interactions appear. In the next subsection we will show that such interactions are induced at the one-loop level when we include the cubic vertices 
of the bulk theory.   One interesting source of boundary interactions is
the field redefinition needed to make all bulk self-interactions of $\zeta$
$O(\epsilon^2)$. As it was 
pointed out in~\cite{m}, all $O(\epsilon)$ bulk cubic interactions
assemble into the form
\beq
S_4^{\zeta^3}= {\epsilon\over 8\pi G}\int d\eta d^3 x f(\zeta)\left [-{d\over d\eta} 
a^2(\eta) {d\over d\eta}\zeta + 
a^2(\eta) \partial^2 \zeta
\right],
\eeq{m31}
where $f(\zeta)$ is a quadratic function of $\zeta$. 
Since they vanish on the linearized equations of motion, in brackets, they 
can be canceled by a local field redefinition of the form 
$\zeta\rightarrow\zeta + f(\zeta)$~\cite{m}. 
When substituted in eq.~(\ref{m26}), this field redefinition generates quartic
boundary interactions
\beq
\Delta S_3^{\zeta^4}={\gamma\over M}{\epsilon\over 8\pi G }
\int d^3 x a(\eta^*)f(\zeta) \partial^2 f(\zeta).
\eeq{m32}
The function $f(\zeta)$ is 
given in~\cite{m}, and we shall use its explicit form in the next subsection, 
to estimate the size of induced boundary interactions. Here, it suffices to 
notice that it contains, among others, terms like 
$(a/2a')^2 (\partial\zeta)^2$, which contain no factors of $\epsilon$. So, 
the interaction~(\ref{m32}) can generate a large boundary mass term through a 
diagram as in figure 4. 

To sum up, small corrections by seemingly benign irrelevant boundary operators
generically induce by radiative corrections large, dangerous, relevant boundary
operators. This problem does not signal an outright inconsistency of non-BD
initial conditions for inflation, but it makes their description in terms
of an EFT unnatural, that is very sensitive to its UV completion.

An equivalent way of stating the problem is simply that the first correction 
to be expected in a generic EFT of initial conditions is 
\beq
\alpha M \int d^3 x a^3(\eta^*)v^2,
\eeq{m33}
where $\alpha$ is a small coefficient at most of order $\epsilon H/M$ (because of 
experimental constraints!). This is the
first universal correction to the BD vacuum that generically dominates over
all others. Any UV modification of physics, be it strings, non-Lorentz
invariant dispersion relations, or simply phase transitions in the late 
stages of inflation, reduces to this same term. In this perspective, 
modifications of the primordial power spectrum seem ill suited to discriminate
among different types of new high-energy physics.

We could have guessed this result because~(\ref{m33}) is the lowest-dimension 
{\em local} operator that can be written with the field $v$ and its 
derivatives. The only question is whether this operator is generated after all.
Our explicit calculation
answered in the affirmative. There is one last subtlety to explain. Locality
depends on the variable we choose to parametrize scalar fluctuations. The
correct one is the canonical scalar field $v$, which has scaling dimension 
one. It is in terms of this field that one finds that radiative corrections
renormalize the coefficients of relevant operators as $M^{3-\Delta}$ etc.
We can rewrite~(\ref{m33}) or any other local operator in $v$ in 
manifestly covariant form using the 3-d metric $h_{ij}$. 
This expression need not be local in $h_{ij}$. One possible covariantization
of~(\ref{m33}) is
\beq
 \alpha M {\epsilon \over 8\pi G}\int d^3 x \sqrt{h} 
{\cal R} \Delta^{-2} {\cal R}, 
\eeq{m33a}  
where $\Delta$ is the covariant scalar Laplacian in 3-d.
\subsection{Naturalness in Inflation III}
The strongest naturalness constraints on the coefficients of the boundary 
irrelevant operator in (eq.~\ref{m26})  arise when we take into account the 
boundary nonlinear self-interactions of the field $\zeta$. 
In this section we  show that the constraints obtained in this way are the 
same as those obtained in ref.~\cite{p2} from backreaction considerations.

In~\cite{m}, the cubic interaction terms of the field $\zeta$ were found to be
\beq
 S^{\zeta^3} =  {\epsilon \over 8\pi G}\int d^3x d\eta a^2(\eta) 
\Big[\epsilon \zeta (\zeta')^2 + f(\zeta) \Box \zeta + \ldots \Big] .
\eeq{f1}
Here $\Box \zeta$ is the term inside brackets in eq.~(\ref{m31}),  
\beq
f(\zeta) = \epsilon \zeta^2 + {a^{-1}(\eta)\over H}\zeta \zeta' + {a^{-2}(\eta)\over H^2}(\partial \zeta)^2 , 
\eeq{f2}
and we have omitted terms of higher order in $\epsilon$.
These cubic interactions give UV divergent Feynman diagrams that induce new 
marginal {\em and} relevant boundary operators of the form
\beq
 O^{0}  = \tilde{\gamma}{\epsilon \over 8\pi G}  \int d^3 x 
a^2(\eta^*)\zeta\zeta', \qquad \qquad O^{1}= \alpha M {\epsilon \over 
8\pi G} \int d^3 x a^3(\eta^*)\zeta^2.
\eeq{f3}
The operator $O^{(0)}$ must be interpreted as explained in section 2.5: 
the $\zeta\zeta'$ term on the boundary is incompatible with a consistent 
variational principle, so it has to be eliminated with 
a field redefinition. Equivalently , if we work only to linear order in 
$\tilde{\gamma}$, it can be transformed into a term of the type $\zeta^2$ by
using  
the unperturbed BD  boundary condition, $\zeta'(\eta^*) = 
\kappa \zeta(\eta^*)$.
 
The same procedure used  in section 2.3 shows that these boundary operators 
lead to  modifications of the primordial power spectrum of the form 
\beq
{\delta P^{(0)}(k)\over P(k)} \sim \tilde{\gamma} g^{(0)}(k \eta^*), 
\qquad \textrm{and} \qquad  {\delta P^{(1)}(k)\over P(k)} \sim \alpha {M\over H} g^{(1)}(k \eta^*),
\eeq{f4}
where $g^{(0)}$ and $g^{(1)}$ are both $O(1)$.\footnote{Explicitly: 
$g^{(0)}(y)= \cos y - y^{-1} \sin y$,  $g^{(1)}(y)= y^{-2}[2 \cos y - 
(1-y^2)y^{-1}\sin 2y]$.} 

\begin{figure}[h]
\centering \epsfxsize=3in \hspace*{0in}\vspace*{.2in}
\epsffile{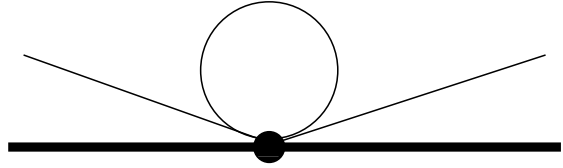} 
\caption{\small Effective boundary two-point vertex induced at one loop by the field redefinition $\zeta \to \zeta + f(\zeta)$. The quartic interaction is proportional to $\gamma f\partial^2 f$.}
\label{selfenergy4}
\end{figure}

The clearest way to exhibit the renormalization effect of the cubic 
interaction on the boundary Lagrangian is to perform  a quadratic field 
redefinition on $\zeta$, of the form  
$\zeta\rightarrow\zeta + f(\zeta)$~\cite{m}. 
This redefinition eliminates from the bulk Lagrangian the  interaction 
proportional to $f(\zeta)$, leaving only cubic terms of $O(\epsilon^2)$ in 
the bulk. On the boundary, on the other hand, the effect of the field redefinition
is to produce new cubic and quartic interactions:
\beq
\int  d^3 x a(\eta^*)\zeta \partial^2 \zeta \;\to\;  \int d^3 x a(\eta^*)\left[\zeta + f(\zeta) \right]\partial^2 \left[\zeta + f(\zeta)\right].  
\eeq{f4a}
Using the explicit form of $f(\zeta)$, eq.~(\ref{f2}), we see that  new 
boundary action contains, among others, the  terms
\beq
\Delta S^b_1 = {\gamma\over M}{\epsilon \over 8\pi G}\int
d^3 x {\epsilon \over H} \zeta^2 \partial^2 \left(\zeta\zeta'\right) , 
\qquad \Delta S^b_2 = {\gamma\over M}{\epsilon \over 8\pi G}\int
d^3 x {1\over a^2(\eta^*) H^3} \zeta\zeta' \partial^2 (\partial\zeta)^2.  
\eeq{f5}
These interactions induce boundary operators of the form (\ref{f3}), 
through the Feynman diagram shown in fig.~\ref{selfenergy4}.

By plugging the interaction $\Delta S^b_1$ in the vertex of this diagram  
we arrive at a contribution to the effective action for $\zeta$:
\bea
\Delta S^{\zeta^2} &&\approx {\epsilon\over 8\pi G }{\gamma\over M} {\epsilon \over H} \int {d^3 q\over (2\pi)^3}\zeta^2(q,\eta^*) \int {d^3 k \over (2\pi)^3}k^2  
\langle \zeta (k,\eta^*)\zeta'(-k,\eta^*) \rangle \\
&& \approx  \epsilon H {\gamma \over M} \int {d^3 q\over (2\pi)^3}\zeta^2 (q,\eta^*) 
\int_{|k|\leq M a(\eta^*)} {d^3 k\over (2\pi)^3} k^2 \eta^{*2} \\ 
&& \approx \epsilon\gamma{M^4 \over  H} \int d^3 x a^3(\eta^*)\zeta^2(\eta^*),
\eea{f6}
where in the third line we have considered only the most divergent term, and 
we used for $\zeta$ the scalar field propagator given in eq.~(\ref{m17}), 
up to an normalization factor $8 \pi G \epsilon^{-1}$:
\beq
\langle \zeta (k,\eta)\zeta(-k,\eta') \rangle = {8\pi G\over\epsilon}\langle 
\chi (k,\eta)\chi(-k,\eta') \rangle.
\eeq{mf6}
The induced boundary term is the operator $O^{(1)}$ in eq.~(\ref{f3}), with a coefficient
\beq
\alpha \sim  \gamma {8 \pi G M^3 \over H} .
\eeq{f7} 

A similar calculation shows that, by plugging the interaction $\Delta S^b_2$ 
in (\ref{f5}) in the vertex of the diagram in fig.~\ref{selfenergy4}, 
one generates 
the operator  $O^{(0)}$ given in eq.~(\ref{f3}), with a coefficient 
\beq
\tilde{\gamma} \sim  {\gamma\over\epsilon} {8 \pi G M^5 \over H^3} .
\eeq{f8} 
By asking that  ${\delta P^{(1)}(k)/P(k)}$ and   ${\delta P^{(0)}(k)/P(k)}$ 
are within the presently acceptable deviation from scale invariance in the 
power spectrum, i.e. that they are smaller that the combination of slow-roll 
parameters $(\epsilon+\eta)$,  we get, from eq.~(\ref{f4}), the constraints 
found in \cite{p2} for the coefficient $\gamma$:
\beq
\gamma \simleq \epsilon  (G M^2)^{-1} {H^2 \over M^2 }, \qquad  
\gamma \simleq \epsilon^2 ( G M^2)^{-1} {H^3\over M^3 }, 
\qquad \gamma \simleq \epsilon \eta (G M^2)^{-1} {H^3\over M^3 }.
\eeq{f9}
Again, we want to stress that these are naturalness bounds, in that they can 
be avoided by tuning the coefficient of the boundary  operators in 
eq.~(\ref{f3}) to a much smaller value than they would generically have in 
the presence of the boundary perturbation, eq.~(\ref{m26}).      
\subsection{Coda: Is the BD Vacuum Natural?}
In the previous subsections we applied standard techniques borrowed from 
field theory to study the naturalness of initial conditions that, below the 
cutoff energy $M$, differ from those specified by the Bunch-Davis vacuum.
One might wonder what happens in the absence of these modifications: 
in particular,   
one might worry that the methods we used make even the unperturbed BD vacuum
fine tuned, thus making the naturalness bounds we found less interesting. 
This would happen if one could find Feynman diagrams producing large 
boundary renormalizations that 
do not arise from the perturbation $\gamma$, i.e. they do not vanish at 
$\gamma=0$.
This is not the case, 
if one uses as the unperturbed vacuum, i.e. the one corresponding 
to $\gamma=0$ in eq.~(\ref{m26}), the vacuum of the {\em interacting} theory.
Indeed, the interacting BD vacuum is defined by the functional integral
\beq
\Psi_{BD}[\zeta]=\int [d\zeta...]\exp(iS),
\eeq{mbd}
where $S=\int d\eta L$ is the action of the {\em interacting} theory, and the 
integration in 
$\eta$ runs from $\eta^*$ to $\eta=-\infty+i\epsilon$. 
The Eucildean continuation in 
$\eta$ selects the BD (or Hartle-Hawking~\cite{hh}) wave function.  
In the language of ref.~\cite{sspds} we are thus choosing ``transparent''  
boundary conditions.
Now, the wave function at any later time $\eta' > \eta^*$ is also
defined by the functional integral~(\ref{mbd}), with the range of integration
ranging from $\eta'$ to $-\infty +i\epsilon$. In this representation, there is 
no boundary at $\eta^*$, so the field redefinition will not produce
any large boundary term.

By splitting the integration in two regions, before and after the ``initial'' 
time $\eta^*$, this obvious result is reinterpereted as a cancelation between
boundary terms arising from the field redefinition of the bulk action
$\int_{\eta>\eta^*}d\eta  L$, and the boundary terms defining the BD vacuum.
So, the only nonlinear interactions left will be bulk 
terms suppressed by powers of 
 the slow-roll parameter. This result shows that the slow-roll inflation 
background is stable 
against both bulk and boundary radiative corrections. 
In the presence of the boundary perturbation~(\ref{m26}), 
however, this argument breaks down, 
and new boundary terms containing $f(\zeta)$ are generated by the field 
redefinition. 

Finally, we should point out again another important feature that 
distinguishes 
the BD vacuum from all others. As we have seen, 
a modification of the BD vacuum by new physics at a physical
energy scale $M$ generically manifests itself in the EFT by the boundary
term~(\ref{m33}). Generically, one would expect the coefficient $\alpha$ to be
$O(1)$, yet observation constrains it to be smaller than $O(\epsilon H/M)$. 
Since this constraint comes from the size of $\delta P (k)/ P(k)$ at 
$|k|=|\eta^*|^{-1}$, i.e. at $|k|_{phys}=|k|/a(\eta^*)=H$, 
an alternative route to make 
the correction~(\ref{m33}) compatible with experiment is to make inflation last
long enough to stretch  $|k|_{phys}$ 
above today's horizon scale. This 
requires $a(\eta_{now})/a(\eta^*)>H/H_{now}$. 
In other words, the change in the initial state of inflation behaves exactly as
any other pre-inflation inhomogeneity. Moreover, its effect is generically
described almost entirely by a single operator, eq.~(\ref{m33}), no matter what
originated the change. So, the BD vacuum is natural also thanks to the very 
property that defines inflation: wait long enough, and all changes will 
be diluted away.

\section{Naturalness in Cyclic Cosmologies}
\subsection{Scale-Invariant Spectrum}
Ref.~\cite{cnz} convincingly showed that a scale-invariant spectrum of 
fluctuations cannot be produced before the bounce in the contracting phase 
described by eq.~(\ref{m7})
with $p \ll 1$. So, it must be created at the bounce. Both effects can be
seen very clearly in our EFT formalism. 

We parametrized the unknown Planckian physics at the bounce in a cyclic 
cosmology in eq.~(\ref{m8}).
By requiring that eq.~(\ref{m8}) is stationary under arbitrary variations in
$\zeta$ and $\Lambda$, including those that do not vanish at the bounce, 
we get the equations of motion for $\zeta$ as well as the junction conditions 
at the bounce 
\bea
\zeta(\eta,x)&\equiv & \int {d^3k \over (2\pi)^3}  \zeta(\eta,k)e^{ikx}, 
\qquad \zeta'' + 2 {b'\over b}\zeta'  + k^2 \zeta=0, \label{m35} \\
b(\eta)&=& {\eta\over\eta^*} \;\;\; \mbox{for} \; \eta \geq \eta^*, \qquad
b(\eta)= \sqrt{2\over p}\left({-\eta\over \eta^*}\right)^{p/(1-p)}     
\;\;\; \mbox{for} \; \eta \leq \eta^*, \nonumber \\
\zeta(\eta^*,k)&=&\alpha(k)\zeta(-\eta^*,k), \qquad
\left.\left({d \over d\eta} \zeta + p \alpha(k)  
\Lambda\right)\right|_{-\eta^*}=0, \label{m36} \\
&&\left.\left({d \over d\eta} \zeta  + F(k)\zeta +
\Lambda\right)\right|_{\eta^*}
= 0 .
\eea{m37}
Before the bounce, the metric can be well approximated by a Minkowski metric, 
for $p\ll 1$, and the general solution to the $\zeta$ equations of motion
is approximately a plane wave
\beq
\zeta(\eta,k)= Ae^{-i|k|(\eta+\eta^*)} + Be^{+i|k|(\eta+\eta^*)}.
\eeq{m37a}
After the bounce, the general exact solution of the equations of motion is
\beq
\zeta(\eta,k)={\eta\over \eta^*}
\left[Ce^{-i|k|(\eta-\eta^*)} + D e^{+i|k|(\eta-\eta^*)}\right], 
\qquad \eta \geq \eta^*.
\eeq{m38}
The wave function of the fluctuation $\zeta(\eta,k)$ is given by the functional
integral
\beq
\Psi[\zeta(\eta,k)]=\int[d\zeta d\Lambda]\exp(iS[\zeta,\Lambda]),
\eeq{m38a}
where $S$ is given in~(\ref{m8}). The boundary condition at 
$\eta\rightarrow-\infty$ is 
$\zeta(\eta,k)\rightarrow Ae^{-i|k|\eta}$, because the
pre-big bang initial state is the standard Minkowski vacuum.
We want to compute the visible primordial spectrum, so we must compute 
$\Psi[\zeta(\eta,k)]$ well after the bounce, at $\eta \rightarrow+\infty$.
So, the value of $\zeta$ at the late-time boundary is fixed: 
$\zeta(\eta,k)\rightarrow \zeta(k)$.
In the quadratic approximation used in~(\ref{m8}), the wave function is 
Gaussian
\beq
\Psi[\zeta(k)]=\prod_k \exp\left[-{\gamma(k)\over 2}\zeta^2(k)\right].
\eeq{m38c}
The power spectrum is a function of the equal time, two point correlator of 
$\zeta$:
\bea
\langle \zeta(k)\zeta(k')\rangle &=& {(2\pi)^3\over k^3}\delta^3(k+k')P(k) 
\label{m38d1}\\  
\langle \zeta(k)\zeta(k')\rangle&=&
\int [d\zeta]\zeta(k)\zeta(k')|\Psi[\zeta]|^2/ \int [d\zeta]|\Psi[\zeta]|^2 
= (2\pi)^3\delta^3(k+k'){1\over 2 \Re \gamma(k)}.
\eea{m38d}

To compute $\gamma(k)$ we perform the functional integral~(\ref{m38a}).
Since it is Gaussian, it reduces to 
\beq
\Psi[\zeta(k)]= \lim_{\eta\rightarrow+\infty}\exp\left[{-i\over 16\pi G}
\int d^3k 
\left({\eta\over\eta^*}\right)^2\zeta'(\eta,k)\zeta(\eta,-k)\right].
\eeq{m38e}
The boundary at $\eta\rightarrow-\infty$ has been discarded using the 
$+i\epsilon$ prescription to select the positive-frequency Minkowski vacuum, 
that is by running the contour of integration in $\eta$ over Euclidean time
in the far past. This prescription selects the true vacuum even in the
interacting theory\footnote{See~\cite{m} for another application of this 
method to cosmology.}. It is this prescription that generates a real
part in $\gamma$, which is naively purely imaginary.

A major simplification occurs because we are interested in observable modes, 
whose wavelength is
much larger than the Hubble radius immediately after the bounce: 
$2\pi/|k| \gg 1/H(\eta^*)= \eta^*$. In this case one can use the boundary 
condition at late time, $\zeta=\zeta(k)$,  to approximate the 
post-bounce evolution of $\zeta$ as
\beq
\zeta(\eta,k)= \zeta(k) + E{\eta^*\over \eta} + O(|k|\eta), \qquad 
{2\pi\over |k|} \gg \eta\geq \eta^*.
\eeq{m38f}
Before the bounce, boundary conditions at $\eta \rightarrow -\infty$ set
\beq
\zeta(\eta,k)= A[1-i|k|(\eta+\eta^*)]+  O(|k|^2\eta^2), 
\qquad  -{2\pi\over |k|} \ll 
\eta \leq -\eta^*.
\eeq{m38g}
The matching conditions at the bounce, eqs.~(\ref{m36},\ref{m37}), 
are then easily solved to give 
\beq
E={{i|k|/ p |\alpha|^2} +F\over {1/ \eta^*} - F -
{i|k|/ p |\alpha|^2}}\zeta(k).
\eeq{m38h}
By substituting into eq.~(\ref{m38e}) we find
\beq
\gamma(k)={1\over 8\pi G}{{|k|/ p |\alpha|^2} -iF\over 1 -\eta^* F -
{i|k|\eta^*/ p |\alpha|^2}},
\eeq{m38i}
whence 
\beq
{(2\pi)^3\over k^3} P(k)= 4\pi G {(1-\eta^*\Re F)^2 + 
(\eta^*\Im F + |k|\eta^*/ p |\alpha|^2)^2
\over |k|/ p |\alpha|^2 -\Im F }.
\eeq{m38j}  

Continuity at the bounce sets $\alpha(k)=1$, $F(k)=0$. In this case we 
get $P(k)\propto |k|^{2} + \eta^* |k|^3 \sim |k|^{2}$, 
instead of the scale invariant 
spectrum $P(k)\propto cst$. This result agrees with ref.~\cite{cnz}. 

This means that {\em all} the features of the primordial spectrum are due to
the unknown physics at the bounce, where the NEC is violated. This is an 
unavoidable~\cite{cnz}, unpleasant feature of cyclic cosmologies, that makes
the whole pre-big bang phase almost useless.
Almost, but not completely. We can consider $\eta^*$ as ``the earliest'' time
and integrate out all the pre-big bang evolution, 
but in doing so, we would need 
nonlocal operators on the brane to generate a scale invariant spectrum.
The long pre-big bang phase, instead, allows for one good 
choice of local coefficients on the brane:
\beq
F(k)=0, \qquad \alpha(k)= {\sigma |k|^2 \eta^{*2}\over\sqrt{p}}, 
\eeq{m38k}
where the constant $\sigma$  can be related to the inflationary parameters 
$H$ and $\epsilon$ introduced in section 3 by comparing eq.~(\ref{m38j}) to the standard inflationary power spectrum, $P(k) = 4\pi G H^2/(2\pi)^3\epsilon $, which leads to $\sigma \eta^* \sim \sqrt{\epsilon}/ H$.

Another natural choice is to require that $\zeta$ is continuous at the
bounce. This sets $\alpha(k)=1$, $F(k)=( \sqrt{p} \sigma \eta^{*2}|k|)^{-1}$, with the same 
constant $\sigma$ as before. We shall call these junction conditions the 
``short bounce.''

An interesting feature shared by all boundary conditions is that $\zeta$
keeps evolving well into the FRW phase, because the coefficient $E$ of the
decaying part of $\zeta$ must be much bigger than $\zeta(k)$ to create the
correct scale-invariant power spectrum. For the junction conditions in 
eq.~(\ref{m38k}), for instance, 
\beq
\zeta_{now}\equiv\zeta(k)\sim\sigma^{-2}(|k|\eta^*)^{-3}\zeta(\eta^*).
\eeq{m38l}
This evolution could be in itself a problem for the cyclic cosmology scenario. 
Eq.~(\ref{m38l}) shows that right after the bounce, in the radiation-dominated
FRW phase, fluctuations are minuscule, compared to today. So, they can be
affected by tiny inhomogeneities developing in the FRW phase, which we have
ignored in our formalism. This is another potential source of fine-tuning, 
besides that due to our ignorance of the physics at the bounce, but
we will not investigate it here.

Finally, the 
post-bounce evolution effectively erases any non-Gaussianity that could
exist at the bounce. We shall discuss this effect in more details in section 6.
\subsection{Causality Concerns}
The parametrization of the unknown Planckian physics at the bounce given in  
eq.~(\ref{m8}) contains instantaneous 
interactions at the bounce time $\pm\eta^*$\footnote{Identification of 
these two times implies an instantaneous bounce. Since we will find acausal 
signals propagating with arbitrarily high speed, this simplification will not
affect our conclusions.}, so acausal propagation of signals 
may occur. To check if this pathology does occur, consider a plane wave 
propagating before the bounce along the direction $x$. The metric before the
bounce is slowly contracting when $p \ll 1$ [see eq.~(\ref{m7})] so we
approximate it by a Minkowski metric as in the previous subsection. 
The wave is then 
\beq
\zeta(\eta,x) = 
\int d\omega e^{-i\omega(\eta+\eta^*-x)}f(\omega), \qquad \eta < -\eta^*.
\eeq{m34}
Before the bounce causality, that is propagation inside the light cone, means
$\zeta(\eta,x)=0$ for $x>\eta+\eta^*+\mbox{constant}$. 
We can set the constant to zero
with a translation of the coordinate $x$. Thus, causality implies that
$f(\omega)$ is analytic in the upper half-plane $\Im \omega > 0$. 

After the bounce, the general exact solution of the equations of motion 
is~(\ref{m38}).

Before the bounce, for the wave given in eq.~(\ref{m34}), $\omega=k$. We choose
$\alpha(\omega)$ real at $\Im \omega =0$. In this
case, the junction conditions reduce to two linear equations for $C,D$
\beq
(1+i|k|) C + (1-i|k|)D -F(k)(C+D) -i{k\over p \alpha^2(k)}(C+D)=0,
\qquad C+D= \alpha(k) f(k).
\eeq{m39}
The coefficient of the right-moving wave is $C$ for $k>0$, and $D$ for $k<0$. 
Since $k=\omega$, this means that such coefficient always 
multiplies the term $\eta \exp(-i\omega\eta)$. Call it $E(\omega)$.
>From eq.~(\ref{m39}) we find 
\beq
E(\omega)=\theta(\omega) C(\omega) + \theta(-\omega) D(\omega) = 
{i\alpha(\omega)\over 2\omega}\left(1-i\omega -F(\omega) - {i\omega\over 
p \alpha^2(\omega)}\right)f(\omega).
\eeq{m40} 
Whenever $F(\omega)$, $\alpha(\omega)$ are analytic in the upper half-plane 
$\Im \omega > 0$, and $\alpha(\omega)$ has no zeroes there, $E(\omega)$ is
analytic in the same region, and no propagation outside the light cone 
occurs after the bounce. This happens for the matching conditions~(\ref{m38k}).
On the other hand, the ``short'' bounce condition, as well as 
most other acceptable ones, is non analytic:
$F(\omega)=1/\sigma \eta^{*2}|\omega|$ has a cut along the $\omega$ 
imaginary axis. 

The bounce explicitly breaks Lorentz invariance, so faster-than light 
propagation does not necessarily imply 
causality violations. After all, one needs a source
before the big bang to get a signal that can propagate faster than 
light {\em after} the big bang. 
Yet, the pathology is troublesome. So, in the
next subsection, we shall restrict out attention to the 
junction conditions~(\ref{m38k}) and see whether they satisfy our naturalness
criterion. Afterward, in section 4.4, we shall return to studying 
more general boundary conditions.
\subsection{Naturalness Concerns I}
The quadratic junction conditions given in subsection 4.1 need to be
covariantized with respect to 3-d coordinate transformations. This is
particularly simple for the conditions~(\ref{m38k}). It suffices to notice
that at linear order $\partial^2\zeta \sim {\cal R}$. Then the action contains 
the term
\beq
S_3={\sigma \eta^{*2}\over 16\pi G\sqrt{p}} \left.\int d^3x \sqrt{h} {\cal R}\Lambda 
\right|_{-\eta^*}.
\eeq{m41}
By expanding this equation we get a coupling 
$S_3^{\zeta^3\Lambda} \sim \zeta^2 \partial^2 \zeta \Lambda$. The equal-time
$\zeta$ propagator is now $\propto |k|^{-1}$ so $S_3^{\zeta^3\Lambda}$ produces
a self-energy diagram as in fig.~3, which gives a one-loop
correction of the form (recall that $a(\eta^*)=1$ here )
\beq
S_3^{\zeta^2} \sim {\sigma \eta^{*2}M^4 \over 4\pi^2\sqrt{p}}\left.
\int d^3x \zeta\Lambda 
\right|_{-\eta^*}.
\eeq{m42}
This correction destroys the scale invariance of the spectrum for all momenta
$|k| < O(\sqrt{GM^4})$. 

Again, a scale invariant spectrum turns out to be possible by fine-tuning 
the operators inserted on the brane. Since they parametrize our ignorance of
the high-energy physics occurring at the bounce, this means that a scale 
invariant spectrum depends on a very special completion of our EFT instead 
of being a robust, generic feature.

Our analysis requires that we do not know in detail the history of the bounce.
If we know what happens at the bounce, then we cannot invoke a genericity 
argument. To appreciate this point, imagine that we ignore the whole history of
the universe before the bounce, i.e. we integrate out $\zeta(\eta)$ for 
$\eta <-\eta^*$ and we substitute for it the boundary term
\beq
S_3={1\over 16\pi G} \left.\int {d^3k\over (2\pi)^3} i |k||\zeta(\eta,k)|^2
\right|_{-\eta^*}.
\eeq{m43}
A generic covariantization of this term would produce a self-energy correction
resulting in a ``mass'' term $\sim M^3 \zeta^2$ that changes the spectrum
for all momenta $|k| \simleq O(GM^3)$. 
In this case though, we {\em do} know the
previous history of the universe. It is a long slowly contracting phase, in
which the power spectrum $1/|k|$ is due to the masslessness of $\zeta$, which
is guaranteed by it being part of a gauge field: the metric.
Essentially the same argument tells us that the scale-invariant 
power spectrum of inflation {\em is} natural.
\subsection{Naturalness Concerns II}
 The function $F(\partial)$ that we introduced in eq.~(\ref{m8}) parametrizes
 how the superhorizon modes get modified during the bounce. As we noted earlier
 in this section, in the case of a short bounce this function must be
 nonlocal to reproduce scale invariance at late times:
\beq
 F(\partial)= \frac{1}{\sqrt{p}\sigma \eta^{*2} \sqrt{-\partial^2} }.
\eeq{jw1}
 The nonlocality of this  interaction on the S-brane
 complicates the study of its stability under radiative corrections.

 Our strategy is to rewrite the part of the action
 containing $F(\partial)$ as a fully local action, by introducing
 auxiliary fields localized on the S-brane. Covariantization of this 
local action introduces then certain universal
 interactions of $\zeta$ with the additional fields, which allow us to 
 study the radiative stability of this system in the same way as in
 the previous section.

\begin{figure}[h]
\centering \epsfxsize=2in \hspace*{0in}\vspace*{.2in}
\epsffile{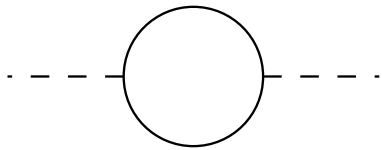} 
\caption{\small The one-loop, UV finite diagram giving rise to the effective action in 
equation (\ref{jw4}). The solid line denotes $\varphi$, whereas dashed lines 
denote $\phi$.} 
\label{phivar}
\end{figure}

 To introduce the technique, consider a three dimensional scalar with
 action:
\beq
 S[\phi]= \int d^3 x \phi \frac{1}{\sqrt{-\partial^2}} \phi.
\eeq{jw2}
 To render this action local, introduce an additional, massless field 
$\varphi$  that couples to $\phi$ as follows:
\beq
 S[\phi,\varphi]= \int d^3 x \left[\lambda \phi \varphi^2+ \varphi \partial^2 \varphi \right].
\eeq{jw3} 
In terms of the variables $\phi,\varphi$, this action is
fully local. The field $\varphi$ appears quadratically in it, so
we can integrate $\varphi$ out exactly. After integration, we are left with an 
effective action for $\phi$ only, which possesses a nonlocal 
propagator. A calculation of the diagram in figure \ref{phivar} gives: 
\beq
  S_{eff}[\phi] = 
\frac{1}{4 \pi} \int d^3 x \phi \frac{\lambda^2}{\sqrt{-\partial^2}} 
\phi + ...,
\eeq{jw4} 
The ellipses denote terms of order $\phi^3$ or higher, that we do not need 
for computing the power spectrum. The quadratic term is indeed the nonlocal 
action of eq.~(\ref{jw2}).
We note that this mechanism depends crucially on the masslessness
of the scalar $\varphi$.  If radiative corrections induce a mass
$m$ for $\varphi$, the form of the action  eq.~(\ref{jw2}) is spoiled for all
momenta $|k| \simleq m$. 

In our case, the fundamental scalar is the $\zeta$-field. We cannot
exactly use the above mechanism with $\phi \to \zeta$, because the
action in eq.~(\ref{jw3}) cannot be covariantized consistently.
Indeed, the interaction term $\zeta\varphi^2$ is uniquely
covariantized by the local expression $\sqrt{h} \varphi^2$. At
zeroth order in the metric fluctuation however, this term gives a
mass to $\varphi$, which makes the one-loop induced term regular in the 
infrared, instead of being divergent as $|k|^{-1}$.

\begin{figure}[h]
\centering \epsfxsize=2in \hspace*{0in}\vspace*{.2in}
\epsffile{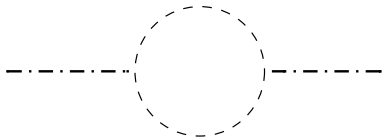} 
\caption{\small The one-loop diagram giving rise to the effective action in 
equation (\ref{jw7}). Dashed lines represent $\psi$, whereas dashed-dotted
 lines represent $\varphi$.} 
\label{phipsi}
\end{figure}

For this reason we must resort to a more complicated system to generate
an action of the form (\ref{jw2}) for $\zeta$. The
following covariant action satisfies our requirements:
\beq
 S[\zeta,\varphi,\psi]= \int d^3 x \sqrt{h}\left[ \alpha {\cal R} \varphi +
 \beta \bar{\psi} ( \Dslash \Delta \varphi) \psi + \bar{\psi} \Delta
  \psi \right].
\eeq{jw5}
 Here $\varphi$ is a scalar,  $\psi$ is a spinor, 
and $\Delta$ is the covariant
 Laplacian. Notice that these fields have non-minimal kinetic terms. 
To see how the effective action for $\zeta$ looks like 
 we proceed in two steps, by first integrating out 
 $\psi$ and next $\varphi$.  Integrating out $\psi$ at  one-loop, as shown in 
figure~\ref{phipsi},  we arrive at an effective action for $\varphi$ and $\zeta$ that to quadratic order reads:
\beq
 S[\zeta,\varphi] =  \int d^3 x \left[ \alpha \partial^2 \zeta \varphi +\beta^2
 \varphi(-\partial^2)^{5/2} \varphi \right].
\eeq{jw6}
Next, integrating out $\varphi$ the effective quadratic action for $\zeta$  reduces to:
\beq
 S_{eff}[\zeta] =  \int d^3 x \left[ \zeta \frac{\alpha^2}{\beta^2
     \sqrt{-\partial^2}} \zeta \right] 
\eeq{jw7}
 By appropriately choosing $\alpha$ and $\beta$, we can
 make $\alpha^2/\beta^2 = (\sqrt{p}\sigma \eta^{*2})^{-1}$, so that this
 interaction at the S-brane is indeed  $\zeta F(\partial) \zeta$.

 The Lagrangian in (\ref{jw5}) is our starting point for
 investigating if the induced action in (\ref{jw7}) is stable under
 radiative corrections. The interactions in (\ref{jw5}) come from
 the covariantization of the action, with $\zeta $ coupling
 universally to the other fields  in the action. Now, we must check if
 any of these interactions feed down to dangerous relevant operators. 
 For us a ``dangerous operator''  means an operator that spoils the scale 
 invariance of the spectrum after the bounce, such as a large mass term for 
 $\psi$ or $\varphi$. In principle, a mass smaller than the IR-cutoff,
 $a H|_{now}$, is allowed, since we cannot probe the spectrum below this scale.
 After all, the function $F(k) \sim |k|^{-1}$ should be regulated by that IR cutoff 
 in order not to generate infinite backreaction at zero momentum. The point is that we will  
 show that the {\it  natural} induced mass is much larger than the IR cutoff. 

\begin{figure}[h]
\centering \epsfxsize=2in \hspace*{0in}\vspace*{.2in}
\epsffile{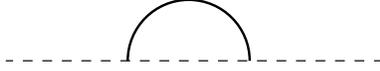} 
\caption{\small The one-loop diagram giving rise to the effective action in 
equation (\ref{jw7}). Dashed lines represent $\psi$ and the solid line
represents $\zeta$.} 
\label{psizeta}
\end{figure}

 For example, we see that the covariant kinetic term for the $\psi$-field
 in eq.~(\ref{jw5}) contains terms like:
\beq
  \Delta S^{\zeta \bar{\psi} \psi} \sim \int d^3x
\bar{\psi} \partial_i \zeta \partial^i \psi.
\eeq{jw8}
 This interaction gives rise to a mass term by the one-loop diagram
 shown in figure \ref{psizeta}. Cutting off the momentum integral at the scale $M$, we obtain:
\beq
 \Delta S^{\bar{\psi} \psi} \sim \frac{M^4}{M_{P}^2} \int d^3x \bar{\psi} \psi.
\eeq{jw9}
 This relevant interaction dominates the original kinetic term 
 for $\psi$ at momentum scales $|k| \simleq {M^2 \over M_p}$. 

 More importantly, 
 the one-loop induced operator in eq.~(\ref{jw6}) is modified to:
\beq
 \Delta S^{\varphi^2} \sim \beta^2 \frac{M_{P}^4}{M^5} \int d^3x
\varphi (\partial^2)^3 \varphi.
\eeq{jw10}
 
 Now, by integrating out the field $\varphi$  we obtain the following
 operator:
\beq
 \Delta S^{\zeta^2} \sim   \frac{1}{\sqrt{p} \sigma \eta^{*2}} \frac{M^5}{ M_{P}^4} 
              \int d^3x \zeta \frac{1}{\partial^2}  \zeta.
\eeq{jw11}
 This term is clearly IR dominant over the original term in
 eq.~(\ref{jw7}), for all momenta $|k| \simleq M (M/M_p)^4$ and it
 produces an unacceptably red spectrum after the bounce over the whole 
 range of scales relevant for observation.
 To avoid the appearance of this term, we must tune the mass term of the 
 fermion in eq.~(\ref{jw9}) to the  much smaller value (by about thirty 
orders of magnitude!) discussed above, by adding a counterterm in the bare 
Lagrangian.  This is 
 the exact analogue of the mass instability of light fundamental scalars 
 in the standard model.
 Our analysis here clearly indicates that the choice for the
 function $F(\partial)$ in eq.~(\ref{jw1}) is fine-tuned in an
 effective field theory perspective, and that only a very specific 
 UV-completion of the short bounce can guarantee scale invariance
 at late times. 

 One can show that similar considerations apply to other choices of boundary non-local operators that reproduce the correct power-spectrum when inserted in eq.~(\ref{m38j}). By an appropriate  choice of auxiliary fields one can make these operators local. This always requires some of the auxiliary fields to have 
an unnaturally small value for the coefficient of an unprotected operator, 
that receives correction due to its universal interactions with the metric.
\section{Non-Gaussianities in Inflation}
\subsection{Minimally Coupled Scalar in de Sitter Space}
A minimally coupled scalar models tensor fluctuations in the CMB, so it has an 
intrinsic interest besides offering a simple example of our method.
The scalar action $S_4+S_3$, given in eq.~(\ref{q1}), is only the quadratic 
part of an action that can contain nonlinear terms both in the 4-d bulk and 
in the 3-d boundary. We shall consider here the effect of adding a boundary 
interaction
\beq
\Delta S_3^{\chi^3}=\int d^3 x a^3(\eta^*) \lambda \chi^3,
\eeq{m44}
while keeping the bulk action $S_4$ quadratic. We could compute the effect of
this term on late-time correlators in a Hamiltonian formalism, as in
subsection 2.3, but we will use instead a Lagrangian approach, which is simpler
for taking into account back-reaction effects and for application to cyclic 
cosmology. In this case, the effect at late
time $\eta\rightarrow 0^-$, is [cfr. eq.~(\ref{m14})]  
\beq
\Psi[\chi_{now}]=\int [d\chi..] e^{i(S_4 + S_3 + \Delta S_3^{\chi^3})}.
\eeq{m45}
The boundary condition at late time is $\chi(0,x)=\chi_{now}(x)$ 
while at early time
is obtained by making the action $S_4+S_3+\Delta S_3^{\chi^3}$ 
stationary w.r.t. free 
variations of $\chi$, including those that do not vanish on the boundary.
The resulting initial condition, written for the Fourier components 
$\chi(0,k)$, is
\beq
\chi'(\eta^*,k) +\tilde{\kappa}(k)\chi(\eta^*,k) + 
3\lambda a(\eta^*)\int {d^3 l\over (2\pi)^3}\chi(\eta^*,l)
\chi(\eta^*,k-l)=0.
\eeq{m46}
In the bulk, $\chi$ obeys a free equation of motion, whose general solution
is
\beq
\chi(\eta,k)= A(1-i|k|\eta)\exp(-i|k|\eta) + B(1+i|k|\eta)\exp(i|k|\eta).
\eeq{m47}
At tree level, the wave function~(\ref{m44}) is  
$\Psi[\chi]=\exp(iS_4+iS_3+i\Delta S_3^{\chi^3})$, 
computed on shell. We only need
quadratic and cubic terms in the action, so we can expand the solution of the
equations of motion as $\chi(\eta,k)=\chi_0(\eta,k) + \lambda \chi_1(\eta,k)$,
and keep only terms at most of linear order in $\lambda$. The coefficients 
$\chi_0(\eta,k),\chi_1(\eta,k)$ obey the boundary conditions
\beq
\chi_0(0,k)=\chi(k), \qquad \chi_1(0,k)=0, \qquad 
\chi'_0(\eta^*,k) +\tilde{\kappa}(k)\chi_0(\eta^*,k)=0.
\eeq{m48}
We need not write the boundary condition for $\chi_1(\eta^*,k)$ because of the
following reason.
On shell and to linear order in $\lambda$, we can integrate by part the free
action and use the free bulk equations of motion to arrive at  
\bea
S_4+S_3+\Delta S_3^{\chi^3} &=& \lim_{\eta \rightarrow 0^-}\left\{
-{1\over 2}\int {d^3k\over (2\pi)^3} a^2(\eta)\chi'_0(\eta,k)
[\chi_0(\eta,-k) + 2\lambda\chi_1(\eta,-k)]+ \right. \nonumber \\ &&  
 {1\over 2}\int {d^3k\over (2\pi)^3} a^2(\eta^*)[\chi'_0(\eta^*,k) 
+\tilde{\kappa}(k)\chi_0(\eta^*,k)][\chi_0(\eta^*,-k)+ 
2\lambda\chi_1(\eta^*,-k)]  + \nonumber \\ && \left. 
\int d^3 x a^3(\eta^*) \lambda \chi_0^3(\eta^*,x)\right\}.
\eea{m49}
Boundary conditions~(\ref{m48}) now make all terms within brackets vanish, 
except the first that reduces to  $\chi(-k)$. Therefore, 
the action reduces to the standard quadratic term, which gives rise to Gaussian
fluctuations, plus a cubic term obtained simply by writing the free field
$\chi$ at $\eta^*$ in terms of its late-time value $\chi(k)$. 
\beq
S_4+S_3+\Delta S_3^{\chi^3}=
\int {d^3k\over (2\pi)^3} i{|k|^3\over H^2}|\chi(k)|^2 + 
\Delta S_3^{\chi^3}[\chi_0^3(\eta^*,k)].
\eeq{m50}
In this formula, we have dropped a real divergent term, which does not 
contribute to expectation values, since it disappears in taking the square
norm of the wave function.
Using eq.~(\ref{m47}) and the boundary conditions~(\ref{m48}), the cubic term 
finally reads
\beq
\Delta S_3^{\chi^3}[\chi_0^3(\eta^*,k)]=\lambda
\int {d^9k\over (2\pi)^9}a^3(\eta^*)(2\pi)^3\delta^3(k_1+k_2+k_3)
\prod_{j=1}^3\chi(k_j)(1-i|k_j|\eta^*)\exp(i|k_j|\eta^*).
\eeq{m51}
The cubic non-Gaussianity is
\beq
 \langle \chi(k_1) \chi(k_2)\chi(k_3)\rangle = {\int [d\chi]\chi(k_1) 
\chi(k_2)\chi(k_3) |\Psi[\chi(k)]|^2\over
\int [d\chi(k)]  |\Psi[\chi(k)]|^2}.
\eeq{m52}
To first order in $\lambda$ this integral is
\bea
\langle \chi(k_1) \chi(k_2)\chi(k_3)\rangle &=& 
6(2\pi)^3\delta^3(k_1+k_2+k_3)\tilde{\lambda}(k_1,k_2,k_3)
\prod_{j=1}^3  {H^2\over 2|k_j|^3} \label{m53} \\
\tilde{\lambda}(k_1,k_2,k_3) &=&  
-2a^3(\eta^*)\Im \left[ \prod_{j=1}^3(1-i|k_j|\eta^*)\exp(i|k_j|\eta^*)\right].
\eea{m54}
Notice that the oscillating exponent in $\tilde{\lambda}$ makes it vanish for
momenta $|k|\gg 1/|\eta^*|$, i.e. for physical momenta $|k|/a \gg H$. So, the
effect of the boundary term is significant only for momenta that crossed
the horizon {\em before} the time $\eta^*$. Physically, this effect can be
understood by interpreting the boundary term $\Delta S^{\chi^3}$ as a 
``summary'' of cosmic evolution prior to $\eta^*$ -such as 
phase transitions, decoupling of heavy particles etc.. In this picture, any
effect of this early history is washed away by inflation in all modes that
keep evolving after $\eta^*$.
This cutoff at $\eta^*$ is another manifestation of the key property of 
inflating backgrounds, namely that they dilute away any primordial 
inhomogeneity. It also makes it difficult to detect such a non-Gaussianity 
unless $Ha(\eta^*)/a(\eta_{now})>H_{now}$.

For momenta $|k| \ll 1/|\eta^*|$ and $\lambda$ real, eq.~(\ref{m54}) becomes
\beq
\langle \chi(k_1) \chi(k_2)\chi(k_3)\rangle = 
-(2\pi)^3\delta^3(k_1+k_2+k_3){H^3\lambda\over 2} \sum_{l>m} 
|k_l|^{-3} |k_m|^{-3}.
\eeq{m55}

\subsubsection{Estimating the Back-Reaction}
The coupling $\lambda \chi^3$ has dimension 3 so it is marginal in the 
boundary EFT. Loops can only renormalize it logarithmically, so it can be 
naturally small. At $O(\lambda^2)$, there exists a potential contribution to 
the boundary ``mass'' term $\int d^3 x a^3 \chi^2$. On dimensional grounds,
we can easily estimate its coefficient as $O(\lambda^2 M) $. So, as long as
\beq
\lambda^2 \simleq \epsilon H/M, 
\eeq{m55a}
the induced ``mass'' term is not in conflict with experimental
bounds.
\subsection{Scalar Fluctuations}
Scalar fluctuations $\zeta$ have model independent 
non-Gaussianities~\cite{m,rm} of the form
\beq
\langle \zeta^3 \rangle =  (2\pi)^3\delta^3(k_1+k_2+k_3){64\pi^2 G^2 H^4\over 
\epsilon} F(k_1,k_2,k_3),
\eeq{m56}
where $F$ is a homogeneous function of degree $-6$ in the momenta.
To compare with the results of the previous subsection, it is convenient to
express the three  point function in terms of the canonically normalized 
field $v$ defined in eq.~(\ref{m19}). Then the size of
$\langle v^3 \rangle$ is $O(\sqrt{8\pi G \epsilon}H^4 k^{-6})$. 
Eventual non-Gaussianities in the initial conditions are instead 
$O(H^3\lambda k^{-6})$. 
Clearly, for $\lambda$ large enough, the 
boundary non-Gaussianity can be dominant for momenta $|k|<1/|\eta^*|$. We must 
only check that the new cubic interactions do not introduce unacceptably 
large boundary ``mass'' renormalization. We already saw that this condition
gives the bound $\lambda \simleq \sqrt{\epsilon H/M}$. 
Another bound comes from the mixed term
due to the universal bulk cubic interaction~\cite{m} 
$\sim \mu v\partial v\partial v $, 
where $\mu=O(\sqrt{8\pi G\epsilon})$. It is a dimension-5
operator that can induce at a boundary term $O(\mu\lambda)$ via the 
self-energy diagram shown in fig. 8. 
\begin{figure}[h]
\centering \epsfxsize=3.5in \hspace*{0in}\vspace*{.2in}
\epsffile{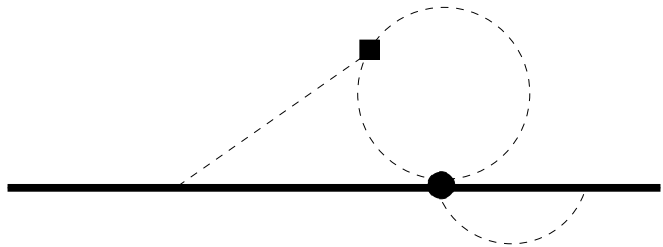} 
\caption{\small A mixed bulk-boundary interaction may generate a boundary mass
term. The black square denotes the bulk interaction 
$\mu v\partial v\partial v $, the
black circle denotes the boundary interaction $\lambda v^3$.} 
\label{selfenergy2}
\end{figure}
On dimensional grounds the induced boundary ``mass'' term is at most 
$O(\lambda\mu M^2) \int d^3 x a^3 \chi^2$. 
This gives another bound on $\lambda$:
\beq
\lambda \simleq 
\sqrt{\epsilon} {H\over M} {M_p\over M}, \qquad M_p \equiv
(8\pi G)^{-1/2}.
\eeq{m57} 
For $\epsilon \sim 10^{-2}$, $M\sim 10^{16}\GeV $ and $H\sim 10^{14} \GeV$   
this bound is even weaker than~(\ref{m55a}).

The ratio of the non-Gaussianity induced by boundary terms 
over the universal bulk term is
$\lambda M_p/\sqrt{\epsilon}H$. For the extremal value 
$\lambda \sim \sqrt{\epsilon H/M}$, it becomes $M_p/\sqrt{HM}=O(10^{4})$. 
So, there is definitely room for an observable signal here! 
A word of caution is necessary, tough. First of all, in order to see a clear
signal we need that inflation does not last too long. Otherwise, thanks to
the cutoff at
$|k| \sim 1/|\eta^*|$, the initial non-Gaussianity 
would only affect unobservable super-horizon fluctuations.
Moreover, in deriving the bound $\lambda \simleq \sqrt{\epsilon H/M}$ we only
demanded that the induced change in the power spectrum is less than 
$\epsilon$. 
More stringent bounds on the power spectrum give a stronger bound on 
$\lambda$~\cite{p2}.
\section{Non-Gaussianities in Cyclic Cosmologies}
\subsection{Damping Effects}
As we mentioned earlier, in cyclic cosmologies, 
a scale invariant power spectrum for scalar fluctuations needs significant
evolution after the bounce, well into the (radiation dominated) FRW phase.
For the junction conditions~(\ref{m38k}), for instance, 
$\zeta_{now}\equiv\zeta(k)=\sigma^{-2}(|k|\eta^*)^{-3}\zeta(\eta^*)= 
\sigma^{-1}(|k|\eta^*)^{-1}\zeta(-\eta^*)$. 
Suppose now that the wave function of the 
fluctuations just before the bounce contains a non-Gaussianity
\beq
\Psi[\zeta_-] = \prod_k \exp\left[-{|k|\over 16\pi G}|\zeta_-(k)|^2 +
\lambda(k_1,k_2,k_3) \zeta_-(k_3)\zeta_-(k_3)\zeta_-(k_3)\right],\qquad
\zeta_-\equiv \zeta(-\eta^*).
\eeq{m100}
 The three-point function can be computed as we did for inflation:
\beq
\langle \zeta_-^3 \rangle = {\int [d\zeta]\zeta^3 |\Psi[\zeta]|^2\over
\int [d\zeta]  |\Psi[\zeta]|^2}\sim \lambda {(16\pi G)^3\over |k|^3}.
\eeq{m101}
Assume for sake of example that bulk non-Gaussianities are negligible; then
the late-time form of $\Psi$ is obtained by evolving $\zeta$ with its quadratic
equations of motion, and by expressing $\zeta_-$ in terms of 
$\zeta_{now}$: $\zeta_-=\sigma |k|\eta^*\zeta_{now}$
\bea
\Psi[\zeta_{now}] &=&\prod_k \exp\left[-{\sigma^2\eta^{*2}|k|^3
\over 16\pi G}|\zeta_{now}(k)|^2 +
\tilde{\lambda}(k_1,k_2,k_3) \zeta_{now}(k_3)\zeta_{now}(k_3)
\zeta_{now}(k_3)\right], \label{m102} \\
\tilde{\lambda}(k_1,k_2,k_3)&=&
\sigma^3\eta^{*3} |k_1||k_2||k_3|\lambda(k_,k_2,k_3).
\eea{m103}
Finally, the magnitude of $\langle \zeta_{now}^3 \rangle$ is
\beq
{\langle \zeta_{now}^3\rangle / (\langle \zeta_{now}^2\rangle)^{3/2}}
\sim (\sigma\eta^* |k|)^3{ \langle \zeta_-^3 \rangle /
(\langle \zeta_-^2 \rangle)^{3/2}} \ll 
{ \langle \zeta_-^3 \rangle /(\langle \zeta_-^2 \rangle)^{3/2}} .
\eeq{m104}
So, any pre-bounce non-Gaussianity is damped by the extremely small factor
$(\sigma\eta^* |k|)^3$. Similar estimates hold for other choices of junction 
conditions.

Evidently, only very dramatic non-Gaussianities have a chance to survive the 
smoothing effect of the bounce. In the next subsection we investigate whether
this may happen.
\subsection{Non-Linear Interactions in the Slow-Contraction Epoch} 
There is one last subtlety about the size of non-Gaussianities in the 
cyclic models we are considering  that needs been addressed.
It relates to the consistency of the perturbative expansion in this 
backgrounds. Looking at the scalar potential (\ref{m7}), one may worry that 
for small $p$ the non-linearities of the theory may become important, and the 
model may become strongly coupled. Indeed, if we consider the 
action to non-linear order, we may worry about terms containing 
self-interactions of the scalar, since on our background, for instance, 
$V'''(\phi) \sim 1/\sqrt{p}$.  

Consider the situation in the  harmonic (de 
Donder) gauge, in which the kinetic term of the metric 
reduces to $\hat h_\mu^\nu\Box \hat h^\mu_\nu$, where 
$\hat h_\mu^\nu= h_\mu^\nu -1/2\delta_\mu^\nu h_\lambda^\lambda$, 
and {\em every} component of the metric propagates. In this gauge,  all 
interactions that involve the background metric  are necessarily  proportional 
to  positive  powers of $H/M_p$ or  $\phi'/M_p$, since we are not 
solving for some of the fields in terms of other fields using the constraints. 
The only potentially large interaction is the cubic term in the expansion for 
the scalar field potential. Using the background solution, eq.~(\ref{m7}), 
this term reads
 \beq
\Delta S^{\phi^3} \simeq \int d^4 x 
{1\over3}\sqrt{{8\pi G\over 3 p}}{1\over \eta^2} \varphi^3 ,
\eeq{fn1}
The crucial fact is that this term is time-dependent. This means that we can trust perturbation theory {\em before} some critical time  $\eta_c$ that will be related to the value of $p$, but not afterwards\footnote{Recall that in the contracting solution we are considering, time runs from $\eta=-\infty$ to $\eta=0$, at which the space-time is singular.}. In other words, we cannot follow the contracting solution arbitrarily close to the singularity.  This means that, if we want to use perturbation theory, we are forced to place our S-brane at an earlier time than $\eta_c$. 

An easy way to get an estimate for $\eta_c$ is to  consider the quartic coupling, $V''''(\phi) \varphi^4$. This is a dimension four operator, and its  dimensionless coupling is $( p M_p^2 \eta^2)^{-1}$. By requiring that this is smaller 
than O(1), we arrive at the estimate $\eta_c \approx (M_p\sqrt{p})^{-1}$. 
Since we know that $\eta^* \sigma= \sqrt{\epsilon} H^{-1} $, and we need
$\eta^*>\eta_c$ our estimate of $\eta^c$ 
translates into $p > (H/\sqrt{\epsilon}M_p)^2 \sigma^2 \approx 10^{-5} \sigma^2$. This should be combined with the requirement that the
initial time is not outside of the effective field theory regime, $\eta^{*}> 1/M$.

\section{Conclusions}
We presented a general method to describe unknown ``dark'' 
periods in cosmology, where high energy/Planckian 
curvature effects become important. These are, among 
others, the Planckian era at the beginning of inflation, and the singularity 
(bounce) in cyclic cosmologies. The unknown cosmic
evolution during such times is replaced by a spacelike brane which can be
endowed with extra degrees of freedom localized on it. These extra
degrees of freedom allow for a general parametrization of the unknown UV 
physics by local operators of ever increasing dimension. 

There is an infinite number of irrelevant operators, which parametrize the 
effect of the dark period and its unknown UV physics, as well as a finite 
number of relevant operators.
On the boundary, relevant operators have dimension strictly less than three.

If the coefficients of the relevant operators are too small, they
may be unstable 
against changes in the UV physics (or, equivalently, the details of the cosmic
evolution during the dark period). The EFT parametrizes our ignorance 
of the dark period, so small coefficients of relevant operators signal 
a non-generic evolution, which is fine-tuned unless we have some 
additional knowledge of the dark period itself. From the point of view of the 
EFT then, such parameters are unnatural.

We applied this naturalness requirement to constrain possible deviations of
the initial state of inflation from the BD vacuum. We found that the generic
signal of any such modification in the power spectrum of primordial 
fluctuations is a universal boundary ``mass'' term, no 
matter what is the origin of the UV modification. From this point of view,
the power spectrum is not
well suited to detect a clear, distinctive signal of new UV physics. For 
instance, trans-Planckian modifications to dispersion relations and late-time
phase transitions during inflation would give an essentially identical signal.

The naturalness requirement is quite powerful in constraining cyclic 
cosmologies. From this point of view, they are all fine-tuned. Many of them 
also allow for faster-than light propagation of signals and may be unstable 
against small inhomogeneities during the radiation-dominated phase of cosmic
evolution. 

We also studied other signals of changes in initial conditions, besides
those affecting the spectrum. We found that a short inflation still allows 
some initial non-Gaussianities to survive and be large enough for 
detection. Unlike the case of changes in the power spectrum, these 
non-Gaussianities can be large without generically inducing a large distortion
of the power spectrum.

Finally, we studied non-Gaussian features in cyclic cosmologies. 
We found that these cosmologies generically damp all pre-bounce 
non-Gaussianities. 
We also showed that the request of no exceedingly large non-linear 
interactions in the pre-bounce, slow contraction epoch, demands that the
S-brane cuts off cosmic evolution at a time parametrically larger than the
Planck time.

Our method extends the formalism set up in~\cite{sspds} to situations where
cosmic evolution is known after {\em and before} a ``dark'' era, but not during
it. The dark era is excised and substituted with an S-brane. It is 
fascinating to speculate that this method may shed new light on other cases
where matter passes through (or ends on) a region that can be traded for an 
S-brane. The most interesting case is of course the singularity inside a 
non-extremal black
hole, which is conjectured to be generically spacelike and of 
BKL~\cite{bkl} type.

We want to end on a constructive note, by pointing out that naturalness could 
also become a guiding principle in the search for a UV completion of cosmic
evolution near singularities, much in the same spirit as naturalness has 
guided particle physics in the search for a UV completion of the standard 
model. 
\subsection*{Acknowledgments}
We would like to thank K. Schalm and J.P. van der Schaar for discussions 
and comments. 
F. Nitti is supported by an NYU McArthur Fellowship.
M.P. is supported in part by NSF grant PHY-0245068, and in part by a Marie 
Curie Chair, contract MEXC-CT-2003-509748 (SAG@SNS). 
J.-W.R. is supported by an NYU Henry~M.~McCracken Fellowship.

\end{document}